\documentclass[a4paper,11pt]{article}
\pdfoutput=1 

\usepackage{jheppub} 

\usepackage[T1]{fontenc} 
\usepackage{xcolor}
\usepackage{subfigure}
\usepackage{stmaryrd}

\newcommand{\beq}{\begin{equation}}
\newcommand{\eeq}{\end{equation}}
\newcommand{\bea}{\begin{eqnarray}}
\newcommand{\eea}{\end{eqnarray}}
\newcommand{\al}{\alpha}

\newcommand{\der}{\partial}

\newcommand{\nn}{\nonumber}
\newcommand{\<}{\langle}
\renewcommand{\>}{\rangle}

\newcommand{\textoverline}[1]{$\overline{\mbox{#1}}$}
\newcommand{\MKK}{M_{\mathrm{KK}}}

\DeclareMathOperator{\Tr}{Tr}

\usepackage{color}

\usepackage[normalem]{ulem}  

\newcommand{\non}{\nonumber\\}

\newcommand{\be}{\begin{equation}}
\newcommand{\ee}{\end{equation}}

\title{\boldmath Heavy Holographic QCD}

\author[a]{Nicolas Kovensky}
\author[a]{and Andreas Schmitt}

\affiliation[a]{Mathematical Sciences and STAG Research Centre, University of Southampton, Southampton
SO17 1BJ, United Kingdom.}

\emailAdd{n.kovensky@soton.ac.uk}
\emailAdd{a.schmitt@soton.ac.uk}

\abstract{We study the phase structure of the Witten-Sakai-Sugimoto model in the plane of temperature and baryon chemical potential, including the effect of a nonzero 
current quark mass. Our study is performed in the decompactified limit of the model, which, at least regarding the chiral phase transition, appears to be closer to real-world QCD than the original version. Following earlier studies, we account for the quark mass in an effective way based on an open Wilson line operator whose expectation value is identified with the chiral condensate. We find that the quark mass stabilizes a configuration with string sources and point out that this phase plays an important role in the phase diagram. Furthermore, we show that the quark mass breaks up the first-order chiral phase transition curve and introduces critical points to the 
phase diagram. Similarities of the phase structure to other holographic approaches and to  lattice simulations of "heavy QCD" are found and discussed. 
By making holographic QCD more realistic, our results open the door to a better understanding of real-world strongly coupled hot and dense matter.
}

\begin{document} 
\maketitle
\flushbottom

\section{Introduction}
\label{sec:intro}

The Witten-Sakai-Sugimoto model \cite{Sakai:2004cn,Sakai:2005yt}, which is based on the gauge/gravity duality \cite{Maldacena:1997re} and in particular on Witten's non-supersymmetric model for low-energy Quantum Chromodynamics (QCD) \cite{Witten:1998zw}, has been successfully applied  to study spectra and properties of mesons, baryons, and glueballs \cite{Sakai:2004cn,Sakai:2005yt,Hata:2007mb,Brunner:2018wbv,Leutgeb:2019lqu}.
It has turned out that the model is not only suited for the physics of single particles in the vacuum, but can also be employed to describe thermodynamic systems and study phase transitions. In its original form, the model is dual to QCD at a large number of colors $N_c$ (at least in a certain limit, which however is inaccessible with current methods). This is borne out in its phase structure \cite{Bergman:2007wp}, which can be mapped to large-$N_c$ QCD predictions \cite{McLerran:2007qj,Philipsen:2019qqm}. In contradiction to real-world, $N_c=3$, QCD, the original version of the Witten-Sakai-Sugimoto model predicts chiral symmetry to remain spontaneously broken for arbitrarily large baryon densities at small temperatures. Approaching the realistic regime of small $N_c$ in a rigorous way is very difficult, although $N_f/N_c$ corrections, where
$N_f$ is the number of flavors, have been discussed \cite{Bigazzi:2014qsa,Li:2016gtz}. A promising alternative is the so-called decompactified limit. In this limit, chiral restoration {\it does} take place at high densities, leading to a more realistic phase structure
\cite{Horigome:2006xu,Li:2015uea}. The price to pay for this improvement is that gluons, while generating the curved geometric background, are decoupled from the dynamics of chiral symmetry breaking. This is not unlike the field-theoretical Nambu--Jona-Lasinio (NJL) model \cite{Nambu:1961tp,Nambu:1961fr}, where gluons are completely absent and similar phase structures have been found \cite{Preis:2012fh}. Examples for applications of the Witten-Sakai-Sugimoto model in that limit to phases of QCD are the discovery of "inverse magnetic catalysis" in dense matter \cite{Preis:2010cq}, and the possibility of a quark-hadron continuity at large chemical potential \cite{BitaghsirFadafan:2018uzs}. These studies have all been performed in the 
chiral limit, i.e., at vanishing {\it current} quark mass (a {\it constituent} quark mass is induced through spontaneous chiral symmetry breaking). In this paper, our  goal is to go beyond the chiral limit and study the effect of an explicit 
chiral symmetry breaking on the phase diagram of the model in the decompactified limit. 

The Witten-Sakai-Sugimoto model is based on a setup of $N_c$ D4-branes and $N_f$ pairs of D8- and $\overline{\rm D8}$-branes. These "flavor branes" represent left- and right-handed fermions and, for $N_c\gg N_f$, can be treated as "probe branes", i.e., they do not backreact on the background geometry determined by the D4-branes. Including a current quark mass in this setup is 
not straightforward because there are no transverse directions that provide a 
massgap for excitations of strings that have endpoints at one color and one flavor brane. Two approaches have been proposed to circumvent this problem, both relying on the physics of open strings connecting the D8- and $\overline{\rm D8}$-branes. In the first approach a bi-fundamental scalar field is introduced, which becomes tachyonic in the infrared and which is used to account for the chiral condensate and the current quark mass  \cite{Bergman:2007pm,Dhar:2007bz,Dhar:2008um}. We follow the second approach, which is based on an open Wilson line between the left- and right-handed flavor branes \cite{Aharony:2008an,Hashimoto:2008sr,McNees:2008km,Argyres:2008sw,Hashimoto:2009hj,Seki:2012tt}. The expectation value of this operator is 
comparable -- but due to its nonlocal nature not identical -- to the usual chiral 
condensate. While this expectation value was calculated and discussed in previous works, it has, to the best of our knowledge, not yet been implemented into a fully consistent calculation of the phase structure, which includes its backreaction on the embedding of the flavor branes. Here we provide such a calculation for nonzero temperatures and chemical potentials. We emphasize that this treatment of the current quark mass is not a rigorous top-down approach. In analogy to ordinary chiral perturbation theory we simply add a term to the action that has the form quark mass times chiral condensate. We shall discuss the novelties in the phase structure of the system together with the conceptual problems that arise in this approach, keeping in mind that results for very large current quark masses have to be treated with care since higher order mass terms are neglected from the beginning. 

Our work is related to various previous studies. Most notably in the holographic context, the phase structure of strongly interacting matter has been studied in the D3-D7 setup \cite{Kobayashi:2006sb}, where a current quark mass is straightforwardly included and the 
interpretation of the chiral condensate is unambiguous \cite{Evans:2010iy,Evans:2011mu,Evans:2011eu}. Despite the differences in how to introduce the quark mass we shall see that both from the holographic point of view as well as in the resulting phase diagrams we find intriguing parallels with the D3-D7 system, which are less striking in the massless limit. Another holographic approach used to address similar questions is a bottom-up approach that works in the Veneziano limit, where the ratio $N_f/N_c$ is held fixed while sending the number of colors to infinity
\cite{Jarvinen:2011qe,Ishii:2019gta}. For example, we shall compute the speed of sound for nonzero temperatures and chemical potentials, which also has been done in that approach \cite{Gursoy:2017wzz}. Last but not least, we shall compare our 
results to recent calculations of QCD on the lattice, where a combined expansion for large coupling and large quark masses is used to circumvent the sign problem and thus is able to provide results for large baryon chemical potentials \cite{Fromm:2011qi,Philipsen:2019qqm}.   

Our paper is organized as follows. In Sec.\ \ref{sec:SSmodel} we introduce 
the model and discuss the mass correction to the action. The solutions to the equations of motion and to the stationarity equations of the free energy density 
are derived in Sec.\ \ref{sec:Solutions} for three different embeddings of the flavor branes. The numerical evaluation with all physical results is presented and discussed in Sec.\ \ref{sec:results}. Or main result is given in Sec.\ \ref{sec:Phases} in the form of phase diagrams for different values of the current quark mass. In Sec.\ \ref{sec:Outlook} we summarize and give an outlook.

\section{Setup}
\label{sec:SSmodel}

\subsection{Brief introduction to the model and Dirac-Born-Infeld action}

We start with a very brief description of the model, to establish the notation and 
the action to which we add the quark mass correction. More detailed 
accounts of the Witten-Sakai-Sugimoto model can be found in the original works \cite{Sakai:2004cn,Sakai:2005yt} and in the review \cite{Rebhan:2014rxa}. 
The model is based on a type-IIA string theory background generated by a large number $N_c$ of D4-branes extended over the directions $(x_0, \vec{x}, x_4)$, where $(x_0, \vec{x})$ are the 4d space-time directions.  
As proposed by Witten \cite{Witten:1998zw}, the $x_4$ direction is compactified 
on a Kaluza-Klein (KK) circle of radius $\MKK^{-1}$ with boundary conditions for the fermions that lead to the breaking of supersymmetry. As we are interested in thermal properties of the system, also the imaginary time direction $\tau = i x_0$ is compactified, with radius $(2\pi T)^{-1}$, where $T$ is the temperature. Thus, while 
the $x_4$ radius is a fixed dimensionful parameter of the model, the $\tau$ radius becomes smaller with increasing temperature. 
The AdS/CFT duality connects the low-energy description of this system with a  non-supersymmetric and non-conformal Yang-Mills field theory. The parameters of the string theory are related to the field-theoretical ones via
\begin{equation}
    \frac{R^3}{\ell_s^{3}} = \pi g_s N_c \, , \qquad \lambda = \frac{2\MKK R^3}{ \ell_s^{2}} \, ,
\end{equation}
where $R$ is the curvature radius of the bulk geometry, $\ell_s$ is the string length, $g_s$ is the string coupling, and $\lambda$ is the 't Hooft coupling. A supergravity description is accurate for small curvature, i.e., for large values of $\lambda$.  
\begin{figure}[t]
    \centering
    \includegraphics[width=\textwidth]{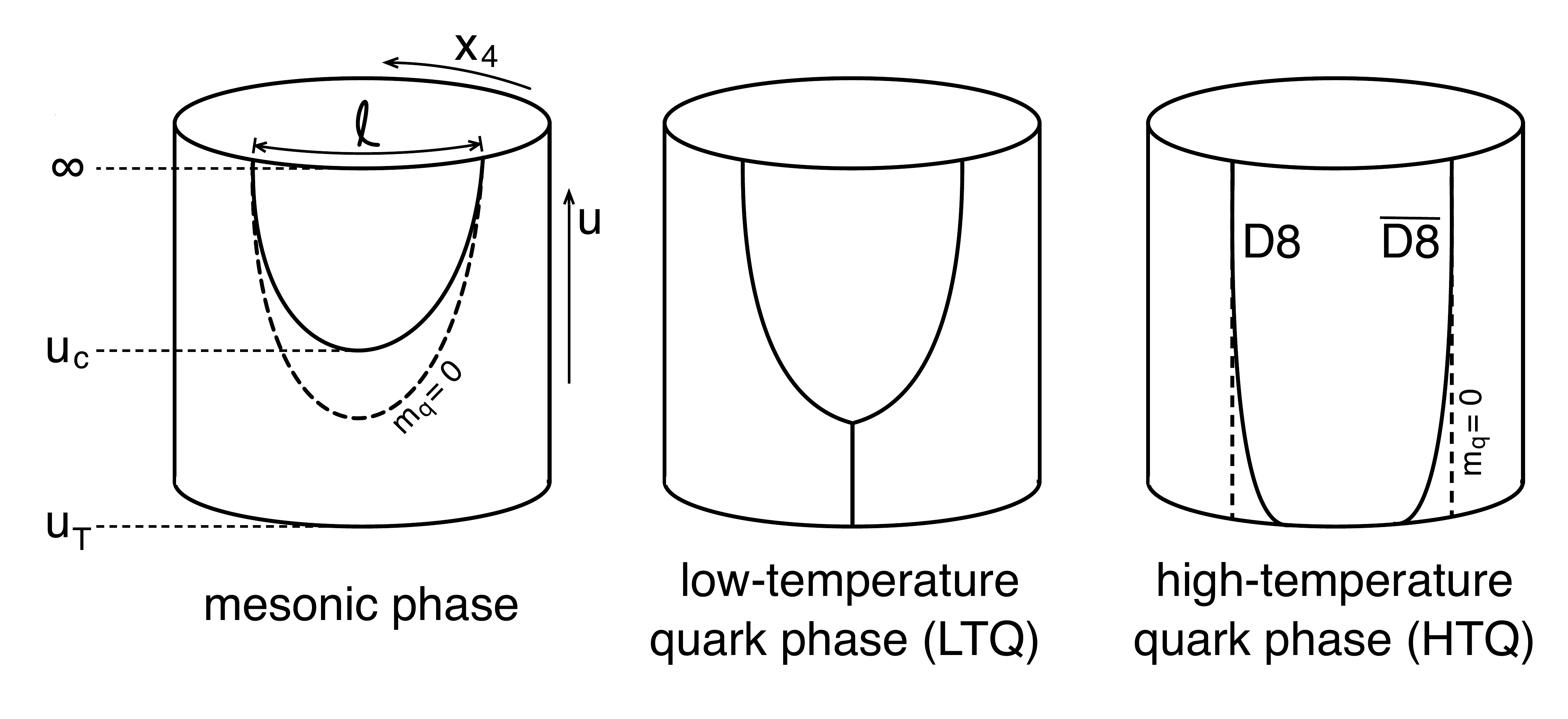}
    \caption{{\small Schematic view of the different embeddings of the flavor branes in the 
    background geometry, here represented by the subspace spanned by $x_4$ and $u$. The asymptotic distance of the flavor branes is given by the dimensionless parameter $\ell$ (such that the $x_4$ radius is 1 and 
    $\ell=\pi$ corresponds to an antipodal embedding). 
   In the mesonic and LTQ phases the flavor branes connect above the horizon $u_T$ at a point that we denote by $u_c$ and which has to be determined dynamically. The mesonic phase has a smooth embedding and vanishing baryon number. In the LTQ phase, which has nonzero baryon number, there is a cusp at which strings are attached, here represented by a straight line. In the HTQ phase, which also has nonzero baryon number, the flavor branes extend all the way to the horizon. The dashed lines indicate the massless limit $m_q=0$ in which there is no stable string configuration and in which the stable phase with nonzero baryon number has straight, disconnected flavor branes. }}
    \label{fig:embeddings}
\end{figure}

At low temperatures, the topology of the manifold spanned by $x_4$ and the radial
("holographic") coordinate $u$ is that of a cigar, while the manifold spanned by 
$\tau$ and $u$ is cylinder-shaped. Above the critical temperature $T_c = \MKK/(2\pi)$ the energetically favored background geometry changes and 
$\tau$ and $x_4$ exchange their roles: a black hole appears.
This Hawking-Page-type transition on the gravity side is usually identified with the deconfinement transition of the gauge theory that lives at the holographic boundary $u=\infty$ (see however \cite{Mandal:2011ws}), and we shall refer to the background where the $\tau$-$u$-manifold is cigar-shaped as the "deconfined geometry". (Strictly speaking, for temperatures larger than the Kaluza-Klein  scale, unwanted 
modes become dynamical and the effective  four-dimensional description breaks down.) In the deconfined geometry, the topology of the 
manifold spanned by $x_4$ and $u$ is a cylinder, as illustrated in 
Fig.\ \ref{fig:embeddings}. This figure also includes the flavor branes (to be discussed momentarily) and their three different stable embeddings in the presence of a quark mass (to be discussed in Sec.\ \ref{sec:Solutions}). 
We will always work in the deconfined background, whose metric is
\begin{equation}
    ds^2 = u^{3/2} \left[f_T(u) d\tau^2 + d\vec{x}^2 + dx_4^2\right] + 
    \frac{1}{u^{3/2}} \left[\frac{du^2}{f_T(u)} + u^2 d\Omega_4^2\right] \, ,    \label{deconf}
\end{equation}
where $d\Omega_4^2$ is the metric of a unit 4-sphere, and the blackening function $f_T$ is 
\begin{equation}
     f_T(u) = 1-\frac{u_T^3}{u^3} \, ,
\end{equation}
where $u_T$ is the position of the horizon in the black hole geometry, related to the dimensionless temperature $t$ via
\begin{equation} \label{tuT}
    t = \frac{3}{4\pi} \sqrt{u_T} \, .
\end{equation}
Here and throughout the paper we work with the same dimensionless quantities used 
before in a series of works \cite{Li:2015uea,Preis:2016fsp,BitaghsirFadafan:2018uzs}. The relation to their dimensionful counterparts involves powers of $M_{\rm KK}$ and the 
curvature radius $R$ as well as some numerical constants, see for instance 
Table I in Ref.\ \cite{Li:2015uea}. Here we restrict ourselves to giving the definitions only of the most important quantities, which are needed to translate our results into physical units. This is done in Table \ref{table1}, where we have introduced the useful abbreviation
\be
\lambda_0 \equiv \frac{\lambda}{4\pi} \, .
\ee

\begin{table*}[t]
\begin{center}
\begin{tabular}{|c|c|c|c|c|} 
\hline
\rule[-0.5ex]{0em}{2ex} 
 temperature & quark chemical  & quark number & 
 constituent & $f_\pi^2m_\pi^2$\\[-2ex]
\rule[-0.5ex]{0em}{2ex} 
& potential & density &
quark mass & \\[1ex] \hline
\rule[-0.5ex]{0em}{5ex} 
$M_{\rm KK} t$ & $\lambda_0 M_{\rm KK}\mu $ & $\;\;\displaystyle{\frac{N_fN_cM_{\rm KK}^3\lambda_0^2}{6\pi^2}n}\;\;$ & 
$\lambda_0 M_{\rm KK}M_q$ & $\displaystyle{\frac{N_cM_{\rm KK}^4e^{\frac{\lambda_0}{\ell}\pi\tan\frac{\pi}{16}}}{3\pi^2} \alpha}$ \\[1ex] \hline
\end{tabular}
\end{center}
\caption{How to obtain the physical, dimensionful quantities from their dimensionless counterparts $t$, $\mu$, $n$, $M_q$, $\alpha$ that are used in the calculation. In all plots in Sec.\ \ref{sec:results} we use an additional rescaling with the asymptotic separation of the flavor branes, $\tilde{t} = t\ell$, $\tilde{\mu} = \mu\ell^2$, $\tilde{n} = n\ell^5$, $\tilde{\Omega} = \Omega\ell^7$, $\tilde{M}_q = M_q\ell^2$, $\tilde{\alpha} = \alpha \ell^4$, $\tilde{\lambda}=\lambda/\ell$. 
}
\label{table1}
\end{table*}
As usual for holographic models,  fundamental matter is introduced by adding probe D-branes. Sakai and Sugimoto included $N_f$ D8-\textoverline{D8} pairs located at the antipodes of the $x_4$ circle. The $U(N_f)_L \times U(N_f)_R$ gauge symmetry on the flavor branes gives rise to chiral symmetry in the field theory. A quark chemical potential is included as the boundary value 
of the temporal component of the abelian gauge field, 
\be \label{boundA}
\hat{a}_0 (u=\infty) = \mu \, ,  
\ee
such that the baryon chemical potential is $N_c\mu$ (we include a hat in the notation of the abelian gauge field for notational consistency 
with previous works).
Below the critical temperature for deconfinement, the topology of the background forces the D8-\textoverline{D8} pair to join in the bulk, forming a $\cup$-shaped configuration. This provides a geometric realization of the spontaneous breaking of chiral symmetry to the diagonal $U(N_f)$. We shall refer to this phase as "mesonic" since the baryon density is zero. Baryons can be added in the form of instantons in the $\cup$-shaped configuration, but here we shall not include this possibility. 
Above the critical temperature, the probe branes are straight and end at the horizon. Since they are disconnected, gauge transformations on D8- and $\overline{\rm D8}$-branes are independent and chiral symmetry remains intact. This phase does have a nonzero baryon density, generated by deconfined quarks. 
%
Chiral restoration and deconfinement occur at the same $\mu$-independent critical temperature in the case of an antipodal separation of the flavor branes. However, this separation $L$ can be 
considered as a second geometric parameter of the model, with $L=\pi/M_{\rm KK}$ corresponding to antipodal separation. We shall work with the dimensionless 
version $\ell = LM_{\rm KK}$, i.e., 
\be\label{boundX4}
x_4(u = \infty) = \frac{\ell}{2} \, , 
\ee
where $x_4(u)$ describes the shape of (half of) the embedding of the probe branes. Therefore, 
the parameters of our model are $\lambda$, $M_{\rm KK}$ and $\ell$. Our main results are of qualitative nature, and thus we do not have to choose specific values for these parameters. Only towards the end of Sec.\ \ref{sec:Phases} we shall briefly discuss some quantitative results using a specific parameter set.

Non-antipodal asymptotic separations, $\ell<\pi$, generalize the model 
\cite{Aharony:2006da} and lead to interesting new results. 
In particular, for sufficiently small $\ell$, namely $\ell < 0.30768\pi$, 
the physics of chiral symmetry breaking and confinement  
get disentangled because the $\cup$-shaped embedding now becomes a solution also in the deconfined geometry, and a free energy comparison between the different embeddings (straight and connected) shows that the chiral phase transition depends on the chemical potential, as expected in real-world $N_c=3$ QCD. We work in the limit $\ell \ll \pi$, where the analysis of chiral symmetry breaking is effectively decoupled from confinement effects,
and the model can be viewed as a holographic version of an NJL-like model
\cite{Preis:2012fh,Antonyan:2006vw,Davis:2007ka}. In particular, we shall work with the deconfined geometry \eqref{deconf} for all temperatures, 
having in mind that by 
decreasing $L$ at given $M_{\rm KK}$ or increasing the $x_4$ radius $M_{\rm KK}^{-1}$  at given $L$ ("decompactification") we can tune the confinement scale to arbitrarily small values compared to the scale for chiral symmetry breaking. Of course, by 
taking this limit we lose some of the top-down control of the original model since now Kaluza-Klein modes become potentially relevant. The gain, however, is a much richer phase structure which has proven to yield interesting physics that are, at the very least, comparable to well-established field-theoretical models
\cite{Preis:2012fh}. It has also been shown that the parameters of the model in the decompactified limit can be fitted to reproduce properties of nuclear matter at saturation density \cite{BitaghsirFadafan:2018uzs}, making it a viable candidate for the study of dense matter in neutron stars. 

In the chiral limit, the different probe brane embeddings and their free energies are obtained from the (Euclidean) Dirac-Born-Infeld (DBI) action 
\begin{equation}
    S_{\mathrm{DBI}} = {\cal{N}}N_f \frac{V}{T} \int_{u_c}^{\infty} 
    du \, u^{5/2} \sqrt{1 + u^3 f_T x_4'^2 - \hat{a}_0'^2} \, ,\label{DBI}
\end{equation}
which is derived from the induced metric on the D8-branes and the dilaton field. Here, 
primes denote derivatives with respect to $u$,  and  we have abbreviated
\begin{equation} \label{N}
{\cal{N}} \equiv \frac{2T_{\mathrm{D}8} V_{S^4}}{g_s} R^5 \left(\MKK R\right)^7= \frac{N_c}{6 \pi^2} \frac{R^2(\MKK R)^7}{(2\pi \al')^3}  = \frac{N_c M_{\rm KK}^4\lambda_0^3}{6\pi^2} \, , 
\end{equation}
where $\al' = \ell_s^2$, $T_{\mathrm{D}8}=(2\pi)^{-8}\ell_s^{-9}$ is the D8-brane tension and $V_{S^4}=8\pi^2/3$ is the volume of the unit four-sphere, while the factor 2 in the first expression accounts for the two halves of the brane worldvolume. If the flavor branes reach the horizon, the lower boundary of the integral in Eq.\ (\ref{DBI}) has to be replaced by $u_T$. We have used translation symmetry in the flat spacetime directions to extract the overall factors of the 3-volume $V$ and inverse temperature. Due to the flavor symmetry, the number of flavors simply appears as a prefactor in front of the action. The grand-canonical potential is given by 
\be\label{Sonshell}
\frac{T}{V} S|_{\mathrm{on-shell}} = {\cal N}N_f\Omega \, , 
\ee
which defines the dimensionless free energy density $\Omega$ as a function 
of the dimensionless thermodynamic variables $\mu$ and $t$. 
The resulting phase diagram in the $\mu$-$t$-plane shows a first-order chiral phase transition \cite{Horigome:2006xu}, separating the chirally broken, $\cup$-shaped
mesonic phase from the chirally restored phase with straight and disconnected branes. We shall discuss how this result is altered by a nonzero current quark mass, which we introduce now.

\subsection{Quark mass correction}

We first recall that, while fundamental quarks are represented by strings stretched between the D4- and the D8-branes, mesons can be described by strings with both endpoints on the D8-branes, and they manifest themselves as scalar and vector fluctuations in the worldvolume theory. When the D8-\textoverline{D8} pairs join in the interior, one finds massless scalar modes in the spectrum. These modes are the Goldstone modes generated by the spontaneous breaking of chiral symmetry. Their fluctuations can be  studied in terms of the chiral field $U(x) = \exp \left[  i \pi(x)/f_\pi \right]\in U(N_f)$, where $x$ stands for the spacetime coordinates and $f_\pi$ is the pion decay constant. For instance, for $N_f=2$,  $\pi(x)$ accounts for the three massless pion modes (and the $\eta$, which however acquires a mass through the axial anomaly). In the holographic theory the chiral field  corresponds to the holonomy 
\begin{equation}
    U(x) = {\cal{P}} \exp \left[  i \int_{-\infty}^{\infty} dz \, a_z(x,z) \right] \, , 
    \label{Udef}
\end{equation}
where $u = (u_c^3 + u_c \, z^2)^{1/3}$, such that the coordinate $z$ parameterizes the entire connected flavor branes, where $a_z$ is the (dimensionless) radial component of the non-abelian gauge field, and where ${\cal{P}}$ denotes path ordering. 
%



In QCD, the pions are of course not massless because chiral symmetry is explicitly broken. In the Witten-Sakai-Sugimoto model, explicit chiral symmetry breaking is  
not as straightforward as in other holographic approaches such as the 
D3-D7 setup. In the latter, one only needs to impose a fixed asymptotic separation between flavor and color branes along one of the transversal directions. The radial dependence of this D3-D7 distance is captured by one of the scalar fields of the worldvolume theory. As usual, once the equations of motion are solved, the corresponding wavefunction allows one to identify a non-normalizable and a normalizable mode, whose coefficients are proportional to the current quark mass $m_q$ and to the expectation value of the quark bilinear, i.e., the chiral condensate  $\<\bar{q} q\>$, respectively.
In the Witten-Sakai-Sugimoto model, however, there is no transverse direction to separate the color from the flavor branes. Moreover, left- and right-handed fermions are separated by the extra dimension $x_4$, such that there is no obvious 
local expression for the chiral condensate. Following Ref.\ \cite{Aharony:2008an}, we consider instead the non-local, but gauge-invariant, operator given by the open Wilson line
    \begin{equation}
        {\cal O}_{ij}(x) = \overline{q}_{i}(x,x_4=-\ell/2)
        {\cal{P}} \exp \left[i\, \int_{-\ell/2}^{\ell/2} dx_4\,  a_4(x,x_4) \right]
        q_{j}(x,x_4=\ell/2) \, , 
    \end{equation}
where $i,j=1,\ldots, N_f$ are flavor indices, and consider the following correction to the action, 
    \begin{equation}
     S_m = - \frac{m_q}{2} \int d\tau d^3x \Tr [{\cal O}+{\cal O}^\dag] \, ,  
    \label{defCFT}
    \end{equation}
where the trace is taken over flavor space, and we have assumed all quark flavors to have the same mass $m_q$. 
This term has the same form as the lowest-order mass term in chiral perturbation theory 
\be\label{MU}
S_m^0 = \int d\tau d^3x \frac{\<\bar{q}q\>_0}{2} \Tr[M(U+U^\dag)] \, , 
\ee
where $M$ is the mass matrix, $M={\rm diag}(m_u,m_d)$ for $N_f=2$, and $U$ is the chiral field, represented by Eq.\ (\ref{Udef}) in the holographic 
theory. The prefactor $\<\bar{q}q\>_0$ is the chiral condensate in the vacuum at vanishing quark mass. With $m_q = (m_u+m_d)/2$ it is related to the pion decay constant $f_\pi$ and the pion mass $m_\pi$ in the vacuum via the Gell-Mann--Oakes--Renner relation \cite{GellMann:1968rz} 
\be \label{GOR}
-m_q\<\bar{q}q\>_0 = \frac{f_\pi^2m_\pi^2}{2} \, .
\ee
The mass term (\ref{MU}) is also used in the approach mentioned in the introduction based on the tachyonic field and moreover was employed (independent of the approach for the chiral condensate) in the study of glueball decay rates in the Witten-Sakai-Sugimoto model \cite{Brunner:2015oga}.

The connection between (\ref{defCFT}) and (\ref{MU}) is made as follows. 
At strong coupling,  the holographic dual of Eq.\ (\ref{defCFT}) can be written in terms of the semi-classical expression for the expectation value $\<{\cal O}\>$. As usual, the open Wilson line corresponds to the Euclidean worldsheet of a fundamental string which ends at the boundary line, and its one-point function is proportional to the exponential of the area of the associated minimal surface, $\<{\cal O}\>\propto e^{-S_{\rm ws}}$. Including boundary terms, we write the worldsheet action as $S_{\mathrm{ws}} = S_{\mathrm{NG}} + S_\der$, where $S_{\mathrm{NG}}$ is the Nambu-Goto action, and the boundary contribution $S_\der = i \int_{-\infty}^{\infty} dz \, a_z$ gives rise to the chiral field (\ref{Udef}) and thus encodes the meson fluctuations. A nontrivial $a_z$ also arises in the presence of baryons, which are instantons on the flavor branes. Therefore, this setup can also be used to compute quark mass corrections to the baryon spectrum \cite{Hashimoto:2009hj,Hashimoto:2009st,Bigazzi:2018cpg}.  
We identify the chiral condensate with the Nambu-Goto part, $\langle{\cal O}\rangle = -\langle \bar{q} q\rangle U$, where $\langle \bar{q} q\rangle = -c e^{-S_{\rm NG}}$
with a proportionality constant $c$ (of mass dimension 3). 
The minus sign is added for convenience such that $c$ is positive.  The chiral condensate in general contains medium and mass corrections and is related to the vacuum chiral condensate at $m_q=0$ by  
\be \label{cond}
\frac{\<\bar{q}q\>}{\<\bar{q}q\>_0} = \frac{e^{-S_{\rm NG}}}{e^{-S_{\rm NG}^0}} \, , 
\ee
where $S_{\rm NG}^0$ is the Nambu-Goto action evaluated in the vacuum and at zero quark mass, which has a simple analytical form, see Eq.\ (\ref{NG0}). We shall use Eq.\ (\ref{cond}) later to compute the normalized chiral condensate.
Using Eqs.\ (\ref{N}), (\ref{GOR}), and assuming the chiral condensate to be uniform, we can write Eq.\ (\ref{defCFT}) as
\bea
    S_m &=&  - \frac{{\cal{N}}}{2} \frac{\alpha}{\lambda_0^3} \, e^{- S_{\mathrm{NG}}} \int d\tau d^3x \, \Tr [
    U + U^\dagger]=   -N_f {\cal{N}}\frac{V}{T} \frac{\alpha}{\lambda_0^3} \, e^{- S_{\mathrm{NG}}}  \, ,
    \label{chiralpert}
\eea
where, in the second step, we have neglected the mesonic fluctuations, i.e., $U=1$, and performed the (then trivial) trace over flavor space and the spacetime integral. 
We have introduced the dimensionless prefactor $\alpha$, which is only nonzero in the presence of a current quark mass $m_q$,
\begin{equation}
\alpha = \frac{cm_q}{N_c}\frac{6\pi^2}{M_{\rm KK}^4} = \frac{3 \pi^2 f_\pi^2 m_\pi^2}{N_c \MKK^4 e^{-S_{\rm NG}^0}} \, . \label{defalpha}
\end{equation} 
In other words, if we write the chiral field $U$ in Eq.\ (\ref{chiralpert})
in terms of the pion field $\pi(x)$, then the prefactor of the quadratic term 
in $\pi(x)$ gives the square of the pion mass. The resulting expression for $m_\pi^2$ is nothing but Eq.\ (\ref{defalpha}). Despite the explicit appearance of $N_c$, the "mass parameter" $\alpha$ does not scale with $N_c$ because $\langle {\cal O}\rangle$ and thus $c$ are proportional to $N_c$ \cite{McNees:2008km}. Notice also that we do not use any explicit form of $c$.
As a consequence, our mass parameter cannot be mapped directly onto the 
current quark mass $m_q$, but only to the combination $f_\pi^2 m_\pi^2 \propto N_c$. We will study how the mass perturbation (\ref{chiralpert}) affects the embedding of the flavor branes for different values of the mass parameter (\ref{defalpha}), while neglecting its effect on the background geometry, which is suppressed by a power of $g_s$ \cite{Aharony:2008an}.  

It remains to compute the Nambu-Goto action. We have
\be
S_{\mathrm{NG}} = 2\lambda_0\int_{u_c}^\infty du \int_0^{x_4(u)} dx_4 \, \sqrt{g}
\, , 
\label{SNG1}
\ee
where the factor 2 accounts for the two halves of the connected flavor branes, 
and the prefactor $\lambda_0$ arises from combining the prefactor $(2\pi \alpha')^{-1}$ that appears in the Nambu-Goto action with the factors from our use of dimensionless coordinates.
With the metric (\ref{deconf}) and introducing an ultraviolet cutoff $\Lambda$, Eq.\  \eqref{SNG1} becomes
\begin{equation}
S_{\mathrm{NG}}
= 2\lambda_0\int_{u_c}^{\Lambda} du \, \frac{x_4(u)}{\sqrt{f_T(u)}} = 2\lambda_0\left[\frac{\ell}{2} \Lambda - \phi_T(u_c)x_4(u_c)  - \int_{u_c}^{\Lambda} du \, \phi_T(u)x_4'(u) \right] \, ,
\label{SNG0}
\end{equation}
where we have integrated by parts, used the boundary condition (\ref{boundX4}), 
and introduced 
\begin{equation}
    \phi_T(u) \equiv \int \frac{du}{\sqrt{f_T(u)}} = \frac{u}{\sqrt{f_T(u)}} \left\{
    1-\frac{3u_T^3}{4 u^3 f_T^{1/6}(u)} {}_2F_1 \left[
    \frac{1}{6},\frac{2}{3},\frac{5}{3}, -\frac{u_T^3}{u^3 f_T(u)}
    \right]\right\} \, , 
\end{equation}
where ${}_2F_1$ denotes the hypergeometric function. For the boundary term we have used $\phi_T(u) \simeq u$ for $u \gg u_T$. At zero temperature,  $\phi_T(u)=u$ for all $u$. 
Later we will also need
\begin{equation} \label{K1}
    \phi_T (u_T) = \frac{3 \sqrt{\pi}}{2} \frac{\Gamma[5/3]}{\Gamma[1/6]} \, u_T 
    \,  . 
\end{equation}
In Eq.\ (\ref{SNG0}) we have allowed for a 
nonzero $x_4(u_c)$, which allows for the flavor branes to be connected with a straight segment constant in $u$. We shall see that this becomes relevant 
for the case where the flavor branes approach the horizon, see Sec.\ \ref{sec:HTQ}. The action (\ref{SNG0}) diverges for $\Lambda\to\infty$. We can simply renormalize it by subtracting the 
vacuum contribution, i.e., by dropping the contribution proportional to $\Lambda$, which does not depend on temperature or chemical potential. In Ref.\ \cite{Seki:2012tt} a temperature-dependent renormalization was employed, which amounts to subtracting the Nambu-Goto action for disconnected, straight branes, where $x_4'=0$ and $u_c$ has to be replaced by $u_T$, 
\be \label{SNGstraight}
S_{\mathrm{NG}}^{||} = 2\lambda_0\left[\frac{\ell}{2} \Lambda - \phi_T(u_T)x_4(u_T) \right] \, .
\ee
Since a theory should be renormalized in the vacuum, before switching on any 
medium effects, this seems incorrect from a theoretical point of view, as argued for instance in Ref.\ \cite{Ewerz:2016zsx} in a similar context, see also footnote 6 in Ref.\ \cite{Bak:2007fk}. The reason why one might be tempted to 
subtract $S_{\mathrm{NG}}^{||}$  is that in this case the renormalized worldsheet action is proportional to the area between the horizon and the holographic boundary that is complementary to the area enclosed by the flavor branes. It is then conceivable to interpret this area as an order parameter 
for chiral symmetry breaking. However, even in this approach the order parameter 
is not proportional to the chiral condensate because the worldsheet action appears in the exponent. (For that reason it was even suggested to interpret $\<{\cal O}\> - 1 $ as an order parameter \cite{Seki:2012tt}, and also the possibility of using $\ln {\cal O}$ was mentioned in a footnote in Ref.\ \cite{Aharony:2008an}.)
Here we follow the theoretically more straightforward way to subtract the vacuum contribution in Eq.\ (\ref{SNG1}), keeping in mind that the interpretation of the chiral condensate has to be taken with care. This might not be too surprising since it is constructed from an open Wilson line operator, which is not identical to a local quark-antiquark operator to begin with. We shall come back to this point when we present the numerical results, see Sec.\ \ref{sec:thermo} and in particular Fig.\ \ref{fig:chiral}. 

To summarize, and to write the mass term in a convenient form for the following  
calculation, we have arrived at
\be \label{Smfinal}
S_m = -{\cal N}N_f\frac{V}{T}\frac{A}{2\lambda_0} \, , 
\ee
where we have abbreviated\footnote{Our choice of defining $A$ without the factor $1/(2\lambda_0)$, which we write explicitly in the action, is convenient, but of course not crucial. It makes the  equations of motion slightly more compact.} 
\begin{equation}
A\equiv \frac{2 \al}{\lambda_0^2} \, e^{- S_{\mathrm{NG}}} \, ,    
\label{Adef1}
\end{equation}
with (after renormalization) 
\be
e^{- S_{\mathrm{NG}}}
= \exp\left\{2\lambda_0\left[\phi_T(u_c)x_4(u_c)  + \int_{u_c}^{\infty} du \, \phi_T(u) x_4'(u) \right]\right\} \, ,
\ee
where the $u$-integral is finite and we have thus reinstated $\Lambda\to \infty$.
While $A$ is constant in the holographic coordinate $u$, it is a functional of $x_4'(u)$ and thus implicitly depends on temperature and chemical potential (it also depends explicitly on temperature). We shall see that via the equations of motion $x_4'(u)$ also depends on $A$ itself, such that Eq.\ (\ref{Adef1}) is an implicit equation for $A$, which we have to solve numerically. The 
mass term (\ref{Smfinal}) introduces -- in addition to the obvious dependence 
on the mass parameter $\alpha$ -- also a nontrivial dependence on the 't Hooft 
coupling. In the absence of a quark mass, $\lambda$ can be formally eliminated by 
an appropriate rescaling of the variables. This is no longer true in the massive case,
and we thus have to discuss the phase structure of the model in the 
two-dimensional parameter space spanned by $\lambda$ and $\alpha$. 

\section{Different probe brane configurations}
\label{sec:Solutions}

In this section we discuss three different probe brane configurations, which appear as classical solutions of the action 
\begin{equation}
S = S_{\mathrm{DBI}} + S_m + S_{\mathrm{sources}} \, .      
\label{Stotal}
\end{equation}
Here, in addition to the DBI action and the mass correction discussed above, 
we have added a source contribution, first considered in Ref.\ \cite{Bergman:2007wp}. The sources are given by static fundamental strings stretching from the tip of the flavor branes all the way down to the horizon (and distributed uniformly along the 3d flat directions), see 
Fig.\ \ref{fig:embeddings} (middle). The endpoints of these strings at $u = u_c$ act as charged sources for the gauge potential $\hat{a}_0(u)$, therefore their number density can be identified with the quark number density\footnote{This is analogous to the point-like baryon source, where an extra contribution to the action is obtained from the Chern-Simons term \cite{Bergman:2007wp}.}. In our conventions, the number of strings $N_s$ is related to the dimensionless baryon number density $n$ by 
\be
n = \frac{6\pi^2}{\lambda_0^2M_{\rm KK}^3N_f N_c }\frac{N_s}{V} \, , 
\ee
and we have 
\begin{equation}
    S_{\mathrm{sources}} = {\cal{N}} N_f \frac{V}{T} n \left[
    (u_c - u_T) - \hat{a}_0(u_c)\right] \, . 
    \label{Sstring}
\end{equation}
Here, the first term is the Nambu-Goto contribution ($u_c-u_T$ simply being the string length), while the second term is the boundary contribution\footnote{In Ref.\ \cite{Bergman:2007wp} the boundary term proportional to $\hat{a}_0(u_c)$ is not written explicitly. We have checked that the Legendre transform performed in this reference leads to the same results. It is somewhat more direct to include this term from the beginning.}. From Eq.\ (\ref{Sstring}) we can 
read off the energy of a single string, which corresponds to the constituent quark mass. We may thus define the dimensionless version of the constituent quark mass, 
in accordance with previous works \cite{Aharony:2008an,Argyres:2008sw}, as 
\be \label{Mq}
M_q = u_c-u_T \, .
\ee
Collecting the constants in Eq.\ (\ref{Sstring}) then gives the dimensionful version, see Table \ref{table1}.

The equations of motion for the abelian gauge field and the embedding function, 
\be
\frac{\delta S}{\delta \hat{a}_0} = \frac{\delta S}{\delta x_4} = 0 \, , 
\ee
become in integrated form
\begin{subequations} \label{EOMs}
\bea
 \frac{u^{5/2}\hat{a}_0'(u)}{\sqrt{1+u^3f_T(u)x_4'^2(u)-\hat{a}_0'^2(u)}} &=& n \, , \label{eom1} \\[2ex]
\frac{u^{5/2}u^3f_T(u)x_4'(u)}{\sqrt{1+u^3f_T(u)x_4'^2(u)-\hat{a}_0'^2(u)}} & = & A\phi_T(u) + k \, , \label{eom2}
\eea
\end{subequations}
with $A$ from Eq.\ \eqref{Adef1} and $k$ being an integration constant. In the presence of string sources, $n$ appears directly due to \eqref{Sstring}; when they are absent, $n$ arises as an integration constant. We can solve Eqs.\ (\ref{EOMs}) 
algebraically for $x_4'$ and $\hat{a}_0'$,
\begin{subequations}\label{solsEOM}
\bea
 \hat{a}_0'(u) &=& \frac{n}{u^{5/2}} \zeta(u)\, , \label{a0prime} \\[2ex]
 x_4'(u) &=& \frac{A \phi_T(u) + k}{u^{11/2}f_T(u)}\zeta(u) \, , \label{x4prime}
\eea 
\end{subequations}
where we have abbreviated 
\begin{equation}
    \zeta(u) \equiv \left[1-\frac{(A \phi_T(u) + k)^2}{u^8 f_T(u)} + \frac{n^2}{u^5}\right]^{-1/2} = \sqrt{1 + u^3 f_T(u) x_4'^2(u) - \hat{a}_0'^2(u)} \, .
\end{equation}
This yields the asymptotic expansions at the holographic boundary
\begin{subequations}
\bea
\hat{a}_0'(u) &=& \frac{n}{u^{5/2}} - \frac{n^3}{2u^{15/2}} + \frac{A^2 n}{2 u^{17/2}} + \cdots \, , \label{a0primeExp}\\[2ex]
  x_4'(u) &=& \frac{A}{u^{9/2}} + \frac{k}{u^{11/2}} + \frac{3 A u_T^3}{4 u^{15/2}} + \cdots  \label{x4primeExp}\, . 
  \eea
\end{subequations}
Via the AdS/CFT dictionary we see that $n$ is indeed the number density associated with the chemical potential $\mu$. Note that the quark mass 
changes the asymptotic behavior of the embedding function $x_4(u)$ since in the presence of a quark mass $A>0$. The boundary conditions at $u=\infty$ are given by Eqs.\ (\ref{boundA}) and (\ref{boundX4}). For the following it is useful to write them in the form 
\begin{subequations} \label{bounds}
\bea
    \mu &=& \int_{u_c}^\infty du \, \hat{a}_0'(u) + \hat{a}_0(u_c) \, ,
    \label{mua0}\\[2ex]
    \frac{\ell}{2} &=& \int_{u_c}^\infty du \, x_4'(u) + x_4(u_c) \, . \label{ell2}
\eea
\end{subequations}
Finally, the dimensionless free energy density, defined in Eq.\ (\ref{Sonshell}), 
is obtained by inserting the solutions \eqref{solsEOM} back into the Euclidean  action. This yields 
\begin{equation} \label{Om}
 \Omega = \int_{u_c}^{\infty} du \, u^{5/2} \zeta(u) - \frac{A}{2\lambda_0} +  n \left[
    u_c - u_T - \hat{a}_0(u_c)\right]. 
\end{equation} 
The free energy depends on the thermodynamic variables $\mu$ and $t$, on the 
externally given parameters $\ell$, $\lambda_0$, $\alpha$ (the dependence on $M_{\rm KK}$ is absorbed in the definition of the dimensionless quantities), and on the variables $k$, $n$, $u_c$, $\hat{a}_0(u_c)$, $x_4(u_c)$, which have to be determined dynamically, depending on the particular boundary conditions we choose at $u=u_c$.

We shall now discuss separately the three phases schematically shown in Fig.\ \ref{fig:embeddings}.

\subsection{$\cup$-shaped embedding: mesonic phase}

First, we consider a $\cup$-shaped embedding, where the branes connect smoothly at $u=u_c$, i.e., we have $x_4(u_c) =0$ and $x_4'(u_c) =\infty$. Then, assuming a 
finite $\hat{a}_0'(u_c)$, Eq.\ (\ref{eom1}) implies $n=0$, which, in turn, with Eq.\ (\ref{a0prime}), gives a constant abelian gauge field, $\hat{a}_0(u) = \mu$. For this solution, the string sources play no role, and so far there is no qualitative difference to the massless case.  We can rewrite the free energy 
(\ref{Om}) as 
\begin{equation}
    \Omega_\cup = \int_{u_c}^{\infty} du \, \frac{u^{5/2}}{\zeta(u)} - \frac{A}{2\lambda_0} + A \int_{u_c}^{\infty} du\, \phi_T(u) x_4'(u)  + \frac{\ell}{2}k \, , 
    \label{OMmesonic}
\end{equation}
where we have used Eqs.\ (\ref{x4prime}), (\ref{ell2}), and $n=0$. This free energy does not depend on $\mu$, which is obvious on physical grounds since the baryon density is zero. The only way to introduce baryon number 
in this phase is through actual baryons, which we ignore throughout this paper. The form \eqref{OMmesonic} is particularly useful for finding the stationary points of $\Omega$ with respect to the remaining variables $k$ and $u_c$. We find that the stationarity condition with respect to $k$ is equivalent to the boundary condition \eqref{ell2}, while the stationarity condition with respect to $u_c$ 
reads
\begin{equation}\label{Omcup}
    0=\frac{\der \Omega_\cup}{\der u_c} = - u_c^{5/2} \sqrt{1-\frac{(A \phi_T(u_c) + k)^2}{u_c^8 f_T(u_c)}} \, .
\end{equation}
Since the square root appears also in the denominator of $x_4'$, see Eq.\ (\ref{x4prime}), stationarity with respect to $u_c$ turns out to be equivalent to the smoothness condition for  the embedding at $u=u_c$, i.e., $x_4'(u_c) = \infty$. 
We use this condition to express $k$ in terms of $u_c$, 
\begin{equation}
 k = u_c^4 \sqrt{f_T(u_c)} - A \phi_T(u_c) \, .
 \label{kms}
\end{equation}
This reduces the evaluation of the mesonic phase to numerically solving 
\begin{equation} \label{ellA}
    \frac{\ell}{2} = \int_{u_c}^\infty du \, x_4'(u) \, ,\qquad  A = \frac{2\al}{\lambda_0^2}
    \exp \left( 2\lambda_0 \int_{u_c}^\infty du\, \phi_T(u) x_4'(u)\right),
\end{equation}
for $A$ and $u_c$ and inserting the solution into the free energy (\ref{OMmesonic}).  

Here and throughout 
the paper we do not expand our equations for small $\alpha$. Such an expansion would be consistent with our starting point, since our quark mass correction 
to the action is linear in $m_q$. However, we have checked that not much is 
gained from this  expansion, the resulting equations are even somewhat more tedious and do not facilitate the numerical evaluation. Therefore, we 
formally keep all orders in the mass, but need to keep in mind that results for very large quark masses can only be obtained as an extrapolation of our approximation.

An analytical result can be obtained from Eqs.\ (\ref{ellA}) in the massless limit at zero temperature. With $A=0$,  $f_T(u)=1$, and $\phi_T(u)=u$ we have with the new integration variable $u'=u/u_c$ (and dropping the prime again)
\begin{subequations}
\bea
\frac{u_c^{1/2}\ell}{2} &=& \int_1^\infty\frac{du}{u^{3/2}\sqrt{u^8-1}} = 2\sqrt{\pi}\, \frac{\Gamma[9/16]}{\Gamma[1/16]} \, , \\[2ex]
\int_{u_c}^\infty du\, \phi_T(u) x_4'(u) &=& u_c^{1/2}\int_1^\infty\frac{du}{u^{1/2}\sqrt{u^8-1}} = \frac{\pi}{2\ell}\tan\frac{\pi}{16} \, ,
\eea
\end{subequations}
where, in the second equation, we have used the result for $u_c$ from the first equation. Consequently, in the vacuum and at zero quark mass we have
\be \label{NG0}
e^{-S_{\rm NG}^0} = \exp\left(\frac{\lambda_0}{\ell}\pi\tan\frac{\pi}{16}\right) \, . 
\ee
This result is identical to that of the confined geometry with non-antipodal flavor branes, see Eq.\ (4.20) of Ref.\ \cite{McNees:2008km}, where this 
result is expressed in terms of the five-dimensional 't Hooft coupling $\lambda_5=2\pi \lambda/M_{\rm KK}$, and where subleading corrections from the Fradkin-Tseytlin contribution were computed additionally (including the factor $N_c$ mentioned below Eq.\ (\ref{defalpha})). 

\subsection{$\Ydown$-shaped embedding: low-temperature quark phase}
\label{sec:LTQ}

Next we consider a $\Ydown$-shaped 
embedding in which the flavor branes connect in a cusp
that is caused by the string sources. This phase has nonzero baryon number, 
but since there are no baryons in the system, baryon number must be created from quarks. Moreover, we shall see that this phase does not exist at large temperatures. Hence we refer to it as the low-temperature quark (LTQ) phase,
see Fig.\ \ref{fig:embeddings} (middle). 
We use the general form of $x_4'$ and $\hat{a}_0'$ from Eqs.\ (\ref{solsEOM}), as well as Eqs.\ (\ref{bounds}) and $x_4(u_c)=0$ to write
the free energy (\ref{Om}) as 
\begin{equation}
    \Omega_\Ydown = \int_{u_c}^{\infty} du \frac{u^{5/2}}{\zeta(u)} - \frac{A}{2\lambda_0} + A \int_{u_c}^{\infty} du\, \phi_T(u)x_4'(u)  + n (u_c - u_T - \mu) + \frac{\ell}{2}k \, .  
    \label{OMstring}
\end{equation}
Again, we need to consider the stationarity equations of the free energy, 
this time with respect to $k$, $u_c$, and $n$. Stationarity with 
respect to all three variables ensures thermodynamic consistency: the 
derivative of $\Omega_\Ydown = \Omega_\Ydown[k(\mu,t),u_c(\mu,t),n(\mu,t),\mu,t]$ with respect to $\mu$ at fixed $t$ is identical to the {\it explicit} derivative if all derivatives with respect to $k$, $u_c$, $n$ are zero. And, as Eq.\ (\ref{OMstring}) shows, the 
explicit derivative is $-n$. Therefore, the identification of $n$ as the baryon density from the AdS/CFT dictionary is in accordance with its thermodynamic 
definition, as it should be. We find that stationarity with 
respect to $k$ is equivalent to Eq.\ (\ref{ell2}), as in the mesonic phase, and thus does not yield any new information. The stationarity equations with respect 
to $n$ and $u_c$ read
\begin{subequations}
\bea
0= \frac{\der \Omega_\Ydown}{\der n} &=&
u_c - u_T - \hat{a}_0(u_c) \, , \\[2ex]
0= \frac{\der \Omega_\Ydown}{\der u_c} &=& 
n -  u_c^{5/2} \sqrt{1-\frac{(A \phi_T(u_c) + k)^2}{u_c^8 f_T(u_c)} + \frac{n^2}{u_c^5}} \, . \label{dOmduc}
\eea
\end{subequations}
The first equation obviously implies 
\begin{equation}
    \hat{a}_0(u_c) = u_c - u_T \, ,  
    \label{a0string}
\end{equation}
which means that $S_{\mathrm{source}}$ does not yield a contribution to the 
free energy (although it affects the solution indirectly through the 
equations of motion).  
From the second relation (\ref{dOmduc}) we derive a condition for $k$ which 
takes 
exactly the same form as in the mesonic case, 
see Eq.\ \eqref{kms}. In particular, the baryon density $n$ does not enter this relation. The only condition for $n$ from Eq.\ (\ref{dOmduc}) is $n>0$, meaning that we always deal with a net baryon, not anti-baryon, density. Inserting the result for $k$ into Eq.\ (\ref{x4prime}) yields
\be \label{x4cusp}
x_4'(u_c) = \frac{u_c}{n\sqrt{f_T(u_c)}} \, , 
\ee
which confirms that there is a cusp at the tip of the connected flavor branes,
$x_4'(u_c)<\infty$, created by the string sources.  

Having eliminated $k$, we are left with solving 
\begin{equation}
     \frac{\ell}{2} = \int_{u_c}^\infty du \, x_4' \, , \qquad  A = \frac{2\al}{\lambda_0^2}
    \exp \left( 2\lambda_0 \int_{u_c}^\infty du\,  \phi_T x_4'\right) 
    \, , \qquad  \int_{u_c}^\infty du \, \hat{a}_0' = \mu - (u_c - u_T)
    \label{Amudefstring}
\end{equation}
for $A$, $u_c$, and $n$ for given $\mu$ and $t$
and inserting the results back into the free energy (\ref{OMstring}). 
In the  limit $n \to 0$, the first two relations are 
identical to those of the mesonic phase and, as Eq.\ (\ref{x4cusp}) shows, the 
brane embedding becomes smooth. This limit is assumed at a certain value of $\mu$ which is given by the third relation: since at zero density we have $\hat{a}'_0(u)=0$, the LTQ phase approaches the mesonic phase at 
the point $\mu = u_c - u_T$, where $u_c$ of course depends implicitly on $\mu$. In other words, at this particular $\mu$ it 
becomes possible to populate the system with quarks. This confirms the 
interpretation of $u_c-u_T$ as the constituent quark mass, as defined in Eq.\ (\ref{Mq}).
The continuous geometrical connection between the mesonic and LTQ phases suggests that a continuous phase transition between them is possible.
We shall see later that such a transition is realized for sufficiently large quark masses. 

The string solution discussed here was already considered in the massless case
\cite{Bergman:2007wp}. It was argued that this phase 
is unstable due to a negative number susceptibility. Our results agree with those of this reference, but we have found that even in the massless limit there is a regime in which the LTQ phase has positive number susceptibility: if the number density in Fig.\ 11 of Ref.\ \cite{Bergman:2007wp} is continued to lower values of $\mu$ one finds a stable branch. Nevertheless, in the absence of a current quark mass, the LTQ phase is metastable at best, i.e., even when it has positive number susceptibility it is energetically disfavored. This is different in the presence of a quark mass, and we shall see that the LTQ phase plays a key role in the phase structure of the model. In fact, we will see that this phase behaves very similarly to a certain solution in the D3-D7 approach \cite{Kobayashi:2006sb, Mateos:2007vc, Evans:2010iy}, where the flavor branes connect to the horizon 
in a "long spike" that "resembles a bundle of strings" \cite{Kobayashi:2006sb}.  The analogy exists not only in the bulk geometry, but will also become manifest by comparing our phase diagrams with the corresponding ones in the D3-D7 model,
see Sec.\ \ref{sec:Phases}. This observation suggests that the string solution considered here might be an approximation to a more complicated, but smooth, embedding of the D8-\textoverline{D8} pair, where the cusp and the string sources are replaced by a spike-like shape of the flavor branes, see also Refs.\  \cite{Callan:1997kz,Gibbons:1997xz}. We leave it to future studies to identify such a solution in the Witten-Sakai-Sugimoto model.

\subsection{$\sqcup$-shaped embedding: high-temperature quark phase}
\label{sec:HTQ}

Finally, we consider a configuration that can be understood as a deformation of the straight-brane configuration in the chiral limit, see Fig.\ \ref{fig:embeddings} (right). This phase also has nonzero baryon number, which 
is created by quarks. 
We shall see that this solution exists only 
for sufficiently large temperatures, and thus we refer to it as the 
high-temperature quark (HTQ) phase\footnote{We thank Josef Leutgeb for pointing out the existence of this phase.}. 
For $m_q = 0$ the brane embedding at non-zero quark density is straight, i.e., in this case $x_4'(u)=0$ due to $k=A=0$, see Eq.\ (\ref{x4prime}). Now, for $m_q>0$ 
and thus $A>0$, $x_4'(u) = 0 $ is not a solution anymore. We still find a solution, however, where the branes extend all the way down to the horizon. 
Therefore, for the HTQ phase, we set $u_c = u_T$ and consider $x_4(u_T)$ as a variable that has to be determined dynamically (for straight branes, $x_4(u_T)=\ell/2$). Then, the (renormalized) Nambu-Goto action  reads 
\begin{equation}
S_{\mathrm{NG}} =  -2\lambda_0 \left[ \phi_T(u_T)x_4(u_T)  + \int_{u_T}^{\infty} du \, \phi_T(u)x_4'(u) \right]\, ,  \label{SNG2}
\end{equation}
where now, as opposed to the previous cases, the lower boundary term is nonzero,
with $\phi_T(u_T)$ being proportional to $u_T$, see Eq.\ (\ref{K1}).
As for the straight-brane solution we require $\hat{a}_0(u_T) =0$. 
Rewriting the free energy (\ref{Om}) with the help of Eqs.\ (\ref{bounds}) 
and replacing the lower boundary of the radial integration by $u_T$ yields
\begin{equation}
    \Omega_\sqcup = \int_{u_T}^{\infty} du \, \frac{u^{5/2}}{\zeta(u)} - \frac{A}{2\lambda_0} + A \int_{u_T}^{\infty} du\, \phi_T(u) x_4'(u)  - n \mu +  k \left[\frac{\ell}{2} - x_4(u_T)\right]. 
    \label{OMquark}
\end{equation}
Now the stationarity conditions with respect to $n$ and $k$ are equivalent 
to the boundary conditions (\ref{mua0}) and (\ref{ell2}), respectively. Hence the
only relevant stationarity equation is 
\begin{equation}
   0 =  \frac{\der \Omega_\sqcup}{\der x_4(u_T)} = - k - A \phi_T(u_T) \, , 
\end{equation}
where the second term originates from the explicit dependence of $A$ on  $x_4(u_T)$. 
We conclude that
\be
k = - A \phi_T(u_T) \, .
\ee
Near the horizon we find
\begin{equation}
    x_4'(u) \simeq \frac{2A}{3 \sqrt{3} u_T^4 \sqrt{1+ \frac{n^2}{u_T^5} - \frac{4 A^2}{9 u_T^6}}} \, \frac{1}{\sqrt{u-u_T}}
    + {\cal O}(\sqrt{u-u_T}) \, ,  
    \label{x4phorizon}
\end{equation}
i.e., the curved flavor branes are horizontal at the horizon, as indicated in Fig.\ \ref{fig:embeddings}. This is in contrast to the massless case, where the straight branes approach the horizon vertically. 

We are left with solving
\begin{equation}
    A = \frac{2\al}{\lambda_0^2}
    \exp \left[ 2\lambda_0\left( \frac{\ell}{2} \phi_T(u_T) +  \int_{u_T}^\infty du\,  \left[\phi_T(u) - \phi_T(u_T) \right] x_4'(u)\right) \right] \, , \quad \int_{u_T}^\infty du \, \hat{a}_0'(u) = \mu
    \label{Adef2}
\end{equation}
for $A$ and $n$, while $x_4(u_T)$ can be computed from \eqref{ell2} afterwards if needed. Of course, only solutions with $x_4(u_T)>0$ make sense, otherwise the branes would cross each other. We find that indeed for small temperatures one runs into solutions with $x_4(u_T)<0$, which have to be discarded. As a result,  there is a ($\mu$-dependent) temperature below which the HTQ phase does not exist.  This is the temperature at which the $\sqcup$-shaped HTQ configuration approaches the $\Ydown$-shaped LTQ solution. At this point $x_4(u_T) = 0$, and it is easy to see that all the conditions we have derived are nothing but the $u_c \to u_T$ limit of those described in the previous subsection. We have thus shown that, on the one hand, the LTQ solution connects continuously to the mesonic solution for $n\to 0$ and, on the other hand, the HTQ solution connects continuously to the LTQ solution for $x_4(u_T)\to 0$. In other words, with the $\Ydown$-shaped string solution we can interpolate continuously between the $\cup$- and $\sqcup$-shaped solutions. This geometric continuity suggests that continuous phase transitions mesonic-LTQ and LTQ-HTQ are conceivable. Whether these transitions are smooth in terms of higher derivatives of the free energy has to be checked numerically, and we will discuss this point in the next section. 

One might ask in which sense the HTQ solution breaks chiral symmetry explicitly. After all, we have introduced a current quark mass, and thus neither the action nor any of the stable states should be chirally symmetric. Geometrically, chiral symmetry in the Witten-Sakai-Sugimoto model is intact (broken) if the flavor branes are disconnected (connected). Now, just as the 
straight-brane solution in the chiral limit, one might think that the HTQ configuration consists of disconnected branes. However, it might be more 
accurate to think of the branes 
in the HTQ phase to be connected by a straight segment at fixed $u=u_T$ from $x_4=-x_4(u_T)$ to $x_4=+x_4(u_T)$ along which $\hat{a}_0(u_T)=0$. This segment gives no contribution to the on-shell action, as can be seen for instance by changing variables to $u(x_4)$, and thus the above results are independent of this 
consideration. The fact that $x_4'(u_T)=\infty$ further suggests that the branes 
"want" to connect, at least the connecting piece can be inserted smoothly.  Connectedness (although not smoothness) is therefore reflected in our choice of the subscript $\sqcup$ for the HTQ phase.  
Another argument for the branes to be connected is the fact
that for disconnected branes the expectation value of the open Wilson line operator has been argued to vanish, $\<{\cal O}\>=0$ \cite{McNees:2008km,Argyres:2008sw}.  Interpreted as an exactly vanishing chiral condensate, this would suggest chiral symmetry to be exactly restored, which, in the presence of a current quark mass, is only expected for asymptotically large temperatures or chemical potentials.

\section{Results}
\label{sec:results}

The remainder of the paper is devoted to the numerical evaluation of the phases just discussed and to the resulting phase diagrams obtained for different values of the quark mass parameter $\alpha$ and the 't Hooft coupling $\lambda$. 
We have reduced the numerical evaluation to solving systems of algebraic equations, which contain numerical integrals. One can easily show that the asymptotic separation of the flavor branes $\ell$ can be eliminated from the equations in each phase by rescaling all quantities with appropriate powers of $\ell$. As a consequence, we will plot only rescaled quantities, which we denote by a tilde. To facilitate translation to the original, un-rescaled quantities, we have collected the definitions of all relevant rescaled quantities in the caption of Table \ref{table1}. This table   therefore shows how to translate the dimensionless, rescaled variables of 
all following results into concrete physical values for given model parameters $\lambda$, $\ell$, $M_{\rm KK}$. 

We
start by discussing the thermodynamics and phase transitions by  fixing either $\mu$ or $t$ in Secs.\ \ref{sec:thermo} and \ref{sec:speed}. Then, we present the phase diagrams in the $t$-$\mu$-plane in Sec.\ \ref{sec:Phases}. Most results are shown for the value of the rescaled 't Hooft coupling $\tilde{\lambda}\equiv \lambda/\ell = 15$, focusing on changes that occur upon variation of the rescaled mass parameter $\tilde{\alpha} \equiv \alpha\ell^4$. The particular value for $\tilde{\lambda}$ is chosen somewhat arbitrarily, but it is almost identical to the value chosen in Ref.\ \cite{BitaghsirFadafan:2018uzs}, obtained from fitting the model parameters to properties of nuclear matter at saturation. (This fit will receive corrections, however, if quark mass effects are included.) At the end we shall also discuss how variations in $\tilde{\lambda}$ affect the phase structure. 

\subsection{Thermodynamics and phase transitions}
\label{sec:thermo}


\begin{figure}[t]
    \centering
      
        \includegraphics[width=0.49\textwidth]{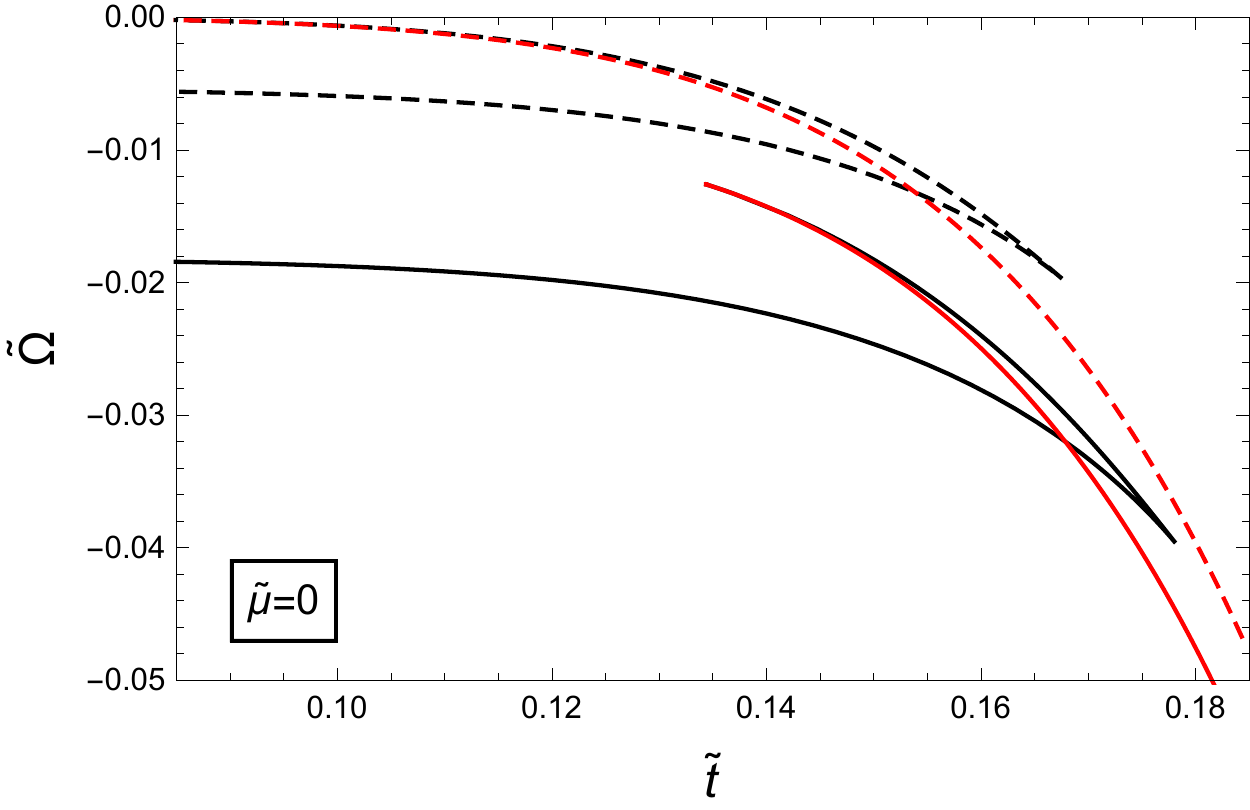}
        \includegraphics[width=0.49\textwidth]{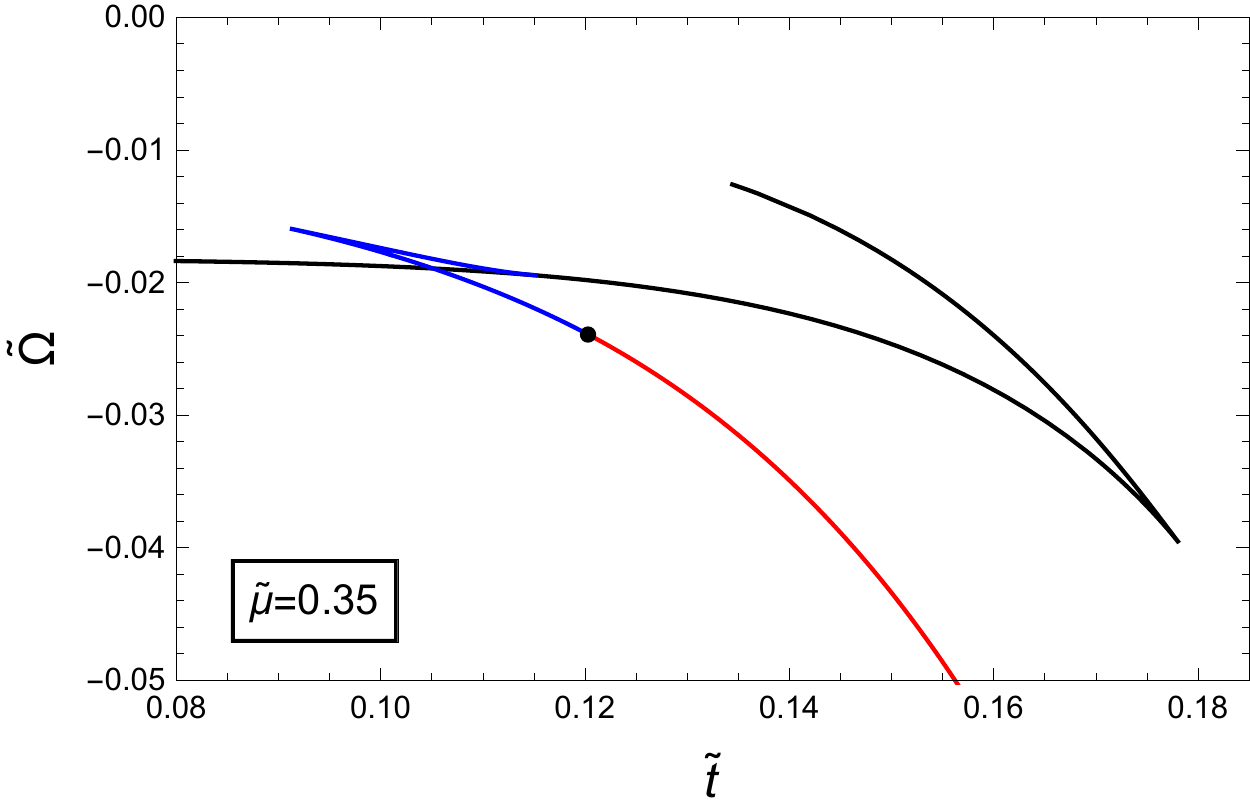}
    
         \caption{{\small Free energy as a function of temperature at two different quark chemical potentials. Solid lines in both panels correspond to a nonzero quark mass, $\tilde{\alpha} = 0.01$, while the 
         dashed curves in the left panel are obtained in the chiral limit $\tilde{\alpha}=0$. Black, blue, and red curves correspond to mesonic, 
         LTQ, and HTQ phases, respectively. This color coding is also used in 
         Figs.\ \ref{fig:chiral} and \ref{fig:NandOMvsMU}. The curve for the mesonic phase in the massive case (right panel) ends at the point where the tip of the connected flavor branes touches the horizon, $u_c=u_T$. Here and throughout Figs.\ \ref{fig:chiral} -- \ref{fig:Phasediagrams}, $\tilde{\lambda} = 15$.}}
         \label{fig:OMvsT}
\end{figure}

\begin{figure}[t]
    \centering
   
        \includegraphics[width=0.49\textwidth]{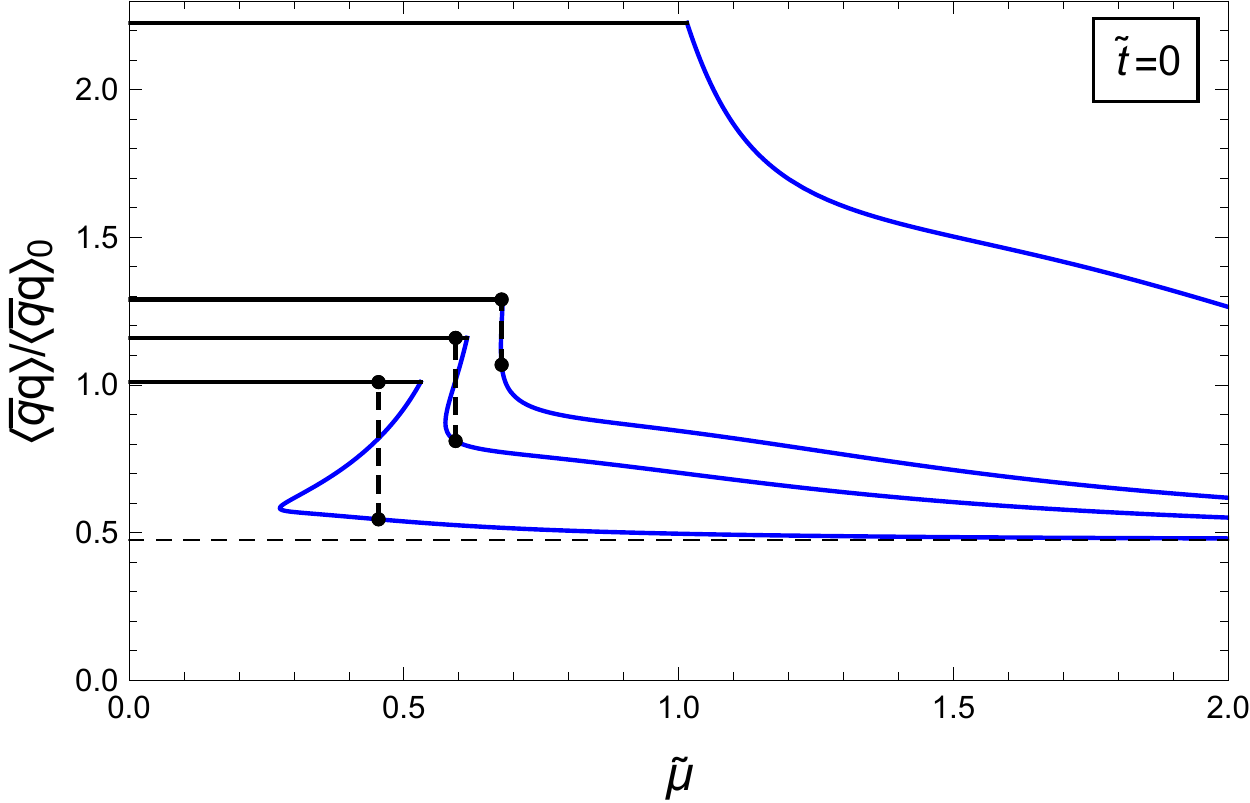}
        \includegraphics[width=0.49\textwidth]{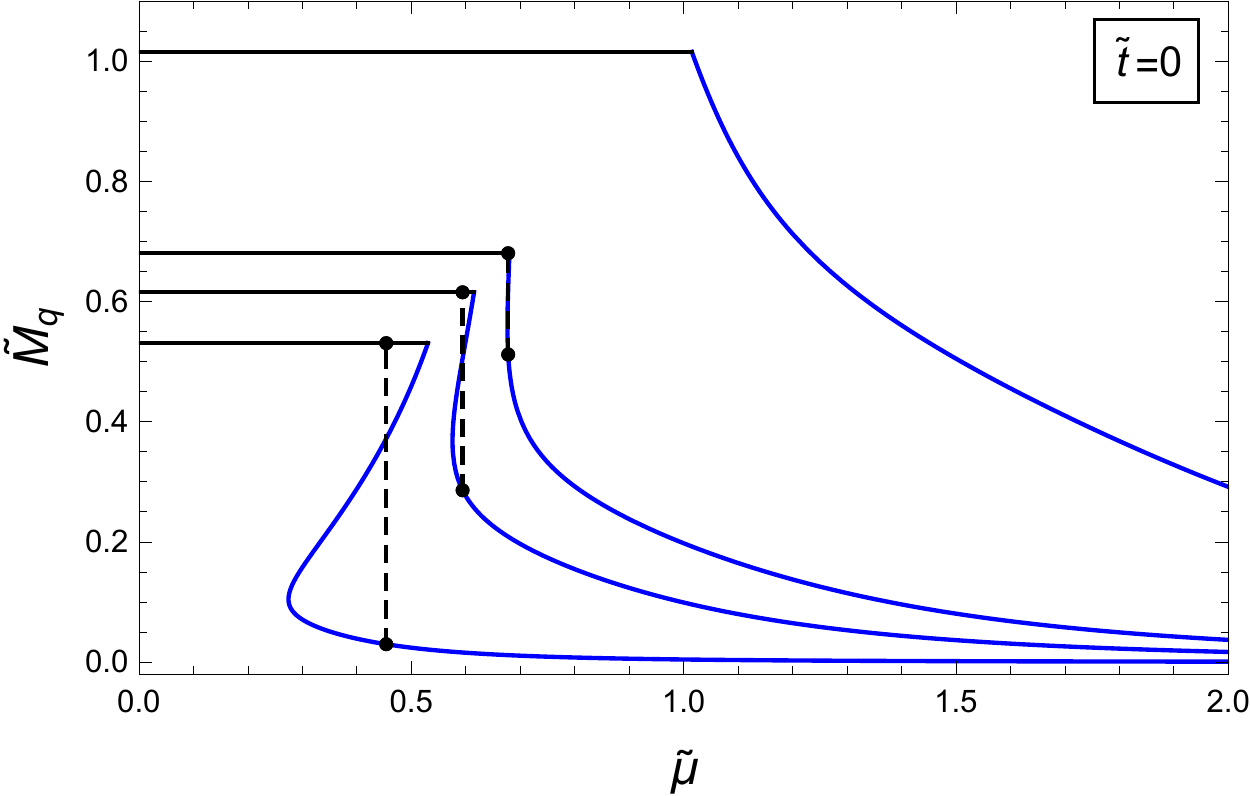}

        \includegraphics[width=0.49\textwidth]{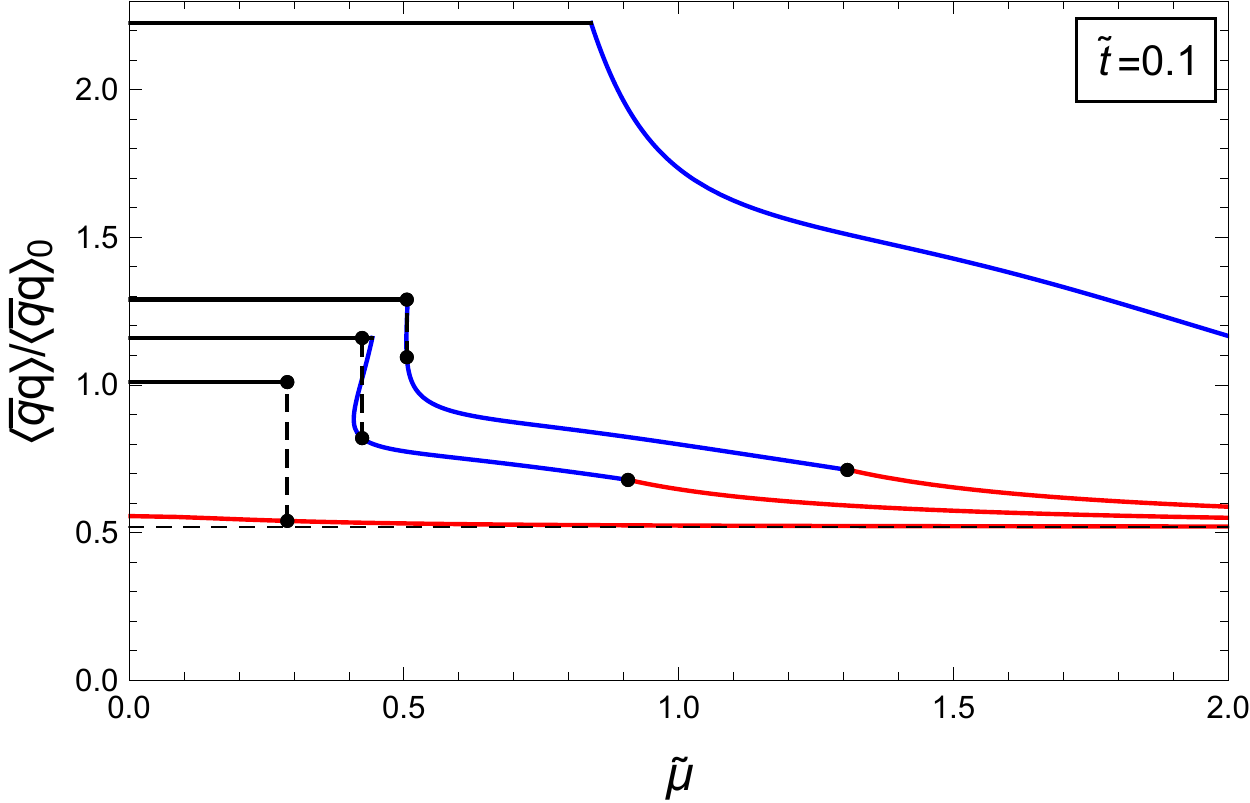}
        \includegraphics[width=0.49\textwidth]{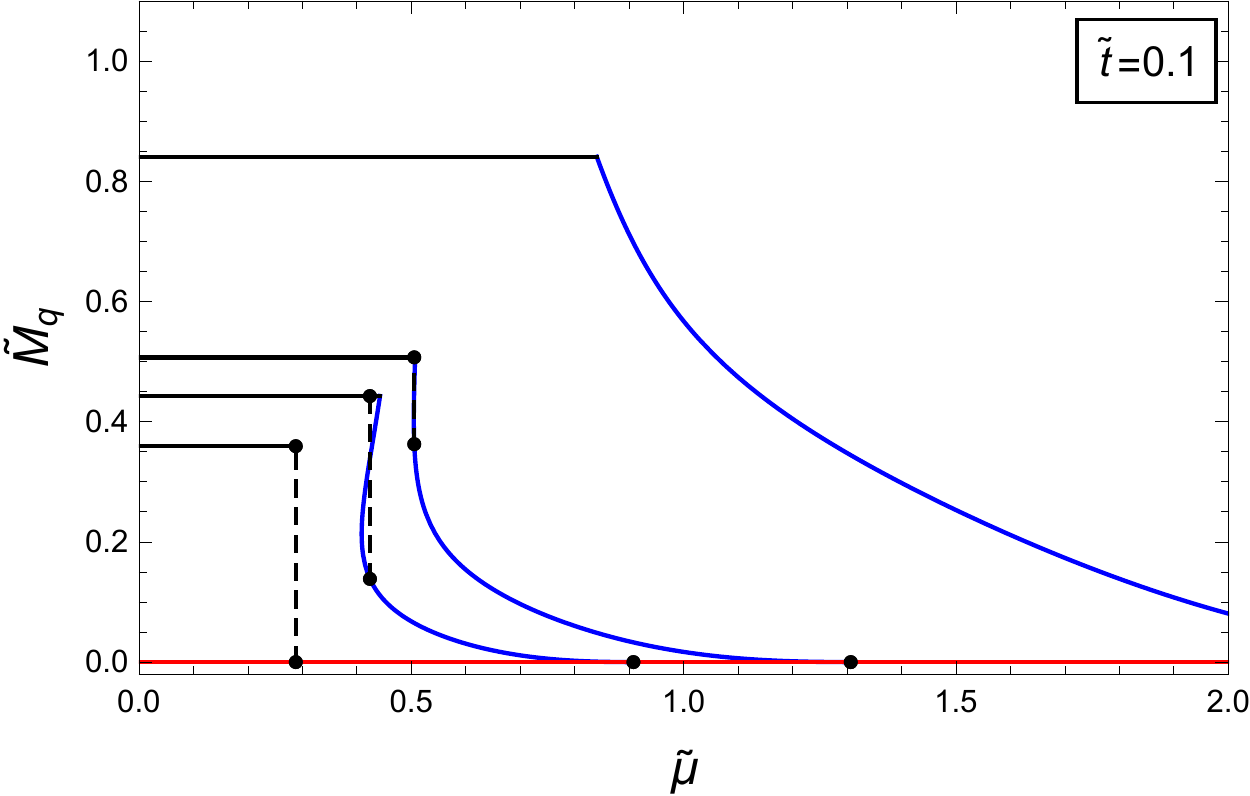}
     
     \caption{{\small Chiral condensate normalized to its vacuum value at zero current quark mass (left panels) and constituent quark mass (right panels) for zero temperature (upper panels) and a nonzero temperature $\tilde{t}=0.1$ (lower panels). In all panels, the mass parameter is $\tilde{\alpha}= 0.001,0.02,0.04,0.2$ from  bottom to top. The vertical dashed segments indicate first-order phase transitions, i.e., some segments of the solid lines are unstable or metastable. The thin horizontal dashed lines in the left panels indicate the 
     asymptotic value (\ref{qqbarasym}) that all curves approach for $\tilde{\mu}\to\infty$.   }}
     \label{fig:chiral}
\end{figure}

\begin{figure}[t]
            \includegraphics[width=0.49\textwidth]{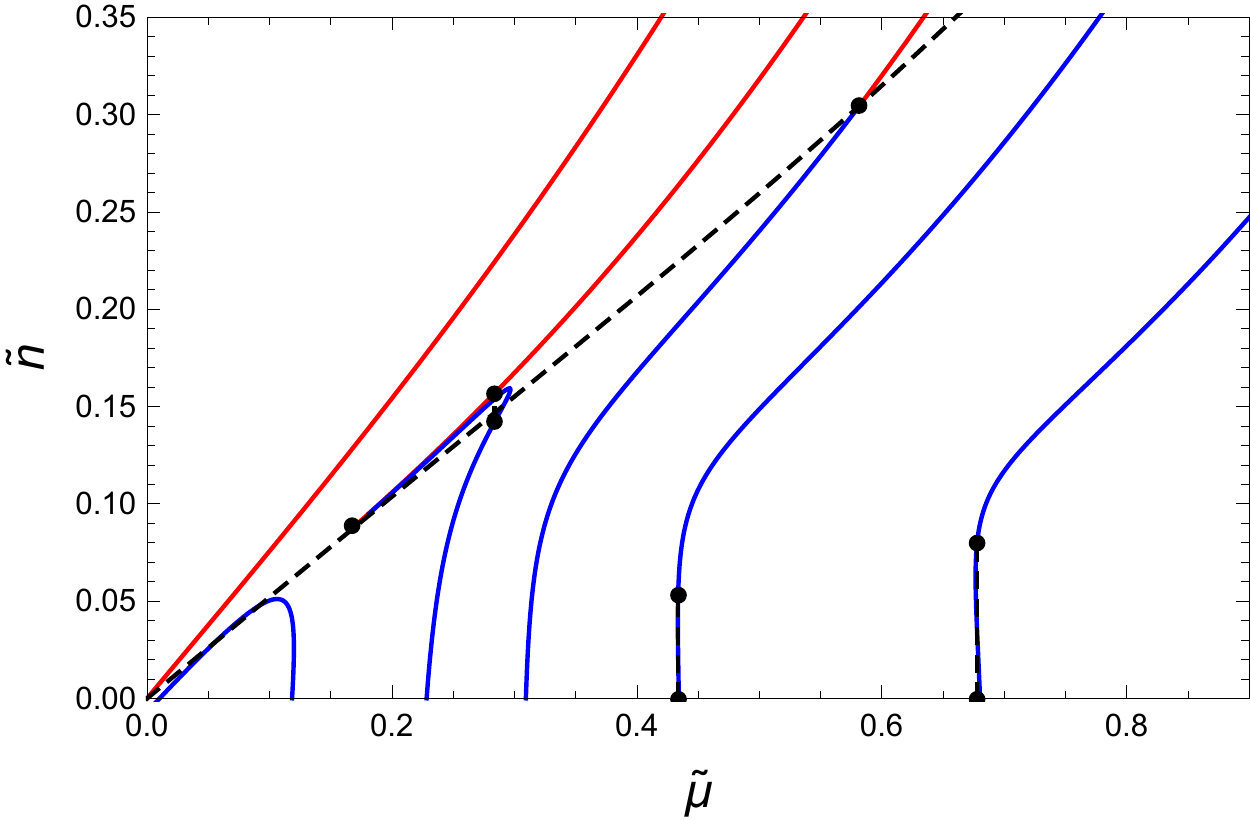}
             \includegraphics[width=0.49\textwidth]{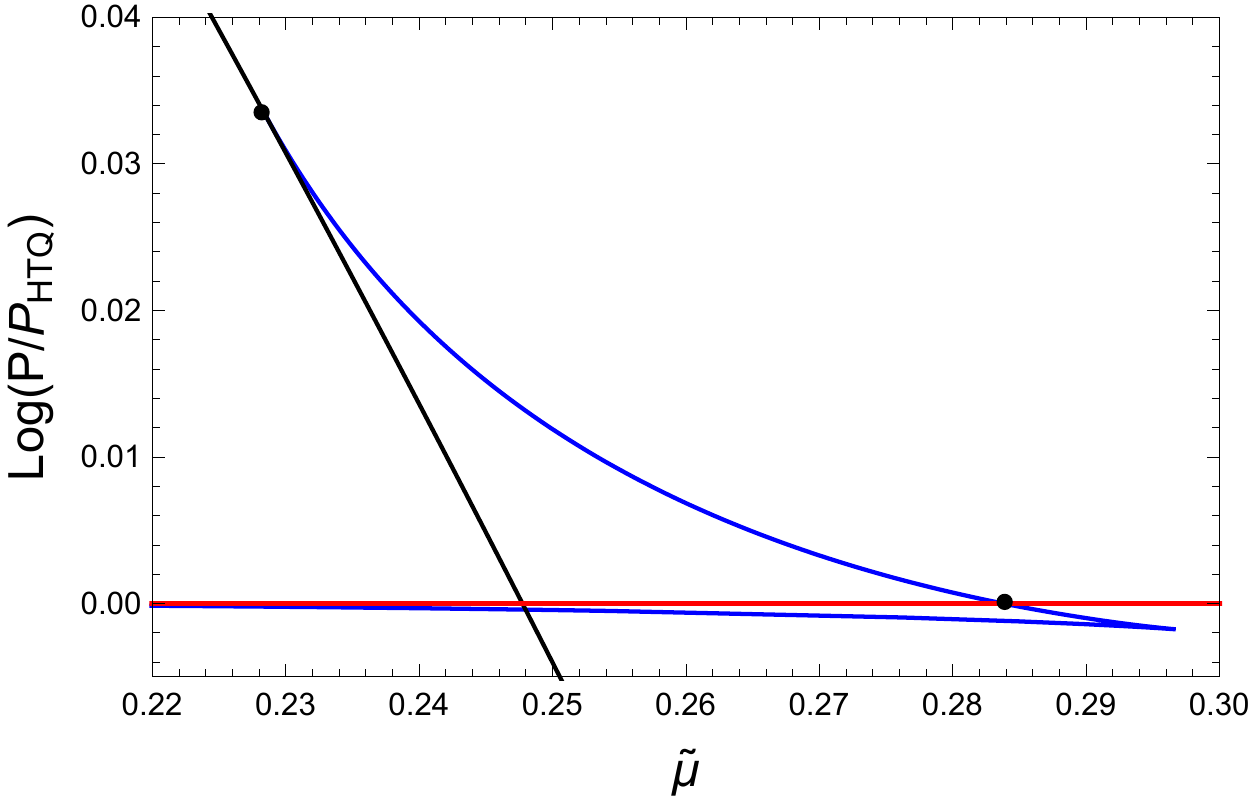}

     \caption{{\small Left panel: Baryon density $\tilde{n}$ as a function of the chemical potential $\tilde{\mu}$ for fixed temperatures $\tilde{t}=0,0.12, 0.15,0.168,0.185$ from right to left. The dashed curve shows where LTQ and HTQ curves connect -- they do so for nonzero but 
     not too large temperatures. The (barely visible) dashed vertical segments for the $\tilde{t}=0,0.12$ curves indicate a discontinuity due to a first-order phase transition. Right panel: Pressure normalized by that of the HTQ phase as a function of $\tilde{\mu}$ for one of the temperatures of the left panel, $\tilde{t}=0.168$, that shows a first-order phase transition between LTQ and HTQ phases. For all curves the mass parameter 
     is $\tilde{\alpha} = 0.04$.}
     }
      \label{fig:NandOMvsMU}
\end{figure}

Figures \ref{fig:OMvsT} -- \ref{fig:NandOMvsMU} collect thermodynamic properties  as a function of either $\mu$ or $t$. They serve to illustrate the structure of the solutions and indicate the phase transitions between them. 

In Fig.\ \ref{fig:OMvsT} we see that in the massless case the mesonic solution 
is two-valued for all temperatures below $\tilde{t} \simeq  0.17$, above which no solution exists. In this case, the straight-brane solution exists for all temperatures and a first-order transition between the two phases occurs. Here, we have shown this solution for $\tilde{\mu}=0$. At larger chemical potentials 
the straight-brane solution, while still existing for all temperatures, moves to lower values of $\tilde{\Omega}$, such that the phase transition moves to lower temperatures until the chirally symmetric phase is preferred for all $\tilde{t}$ if $\tilde{\mu}$ is sufficiently large. This results in the phase diagram of the model in the decompactified limit at zero current quark masses, which is well known \cite{Horigome:2006xu}.

For nonzero quark mass, the mesonic solution becomes single-valued for small 
temperatures since the unstable branch does not reach back all the way to zero temperature, i.e., the point where the tip of the connected flavor branes touches the horizon is now at nonzero temperature. At vanishing chemical potential (left panel), not much is changed compared to the massless scenario: the HTQ phase connects to the unstable branch of the mesonic phase and again there is a first-order transition between the two phases. We have checked that the discontinuity of this transition becomes smaller as we increase the quark mass parameter, but a continuous transition is never obtained. In fact, if $\tilde{\alpha}$ is chosen too large, the mesonic solution ceases to exist even for small temperatures. This would imply that there is no stable phase in a certain region of the phase diagram, which is unphysical. As discussed above, we cannot trust our approximation to arbitrarily large masses, and thus this 
behavior is not surprising. We shall therefore only consider mass parameters  $\tilde{\alpha}$ -- more precisely, only parameter pairs $(\tilde{\lambda},\tilde{\alpha}$) -- for 
which we can determine a stable equilibrium state for all $\tilde{t}$ and $\tilde{\mu}$. 

A more interesting situation, qualitatively different from the massless case,  occurs for nonzero chemical potentials, as shown in the right panel of Fig.\ \ref{fig:OMvsT}. Here, the mesonic and HTQ solutions do not connect continuously anymore. For sufficiently large $\tilde{\mu}$ we find that at the endpoint of the  HTQ solution this solution has smaller free energy than the mesonic solution. Without a third solution, this would imply an unphysical jump in the free energy. This third solution is the string solution: it connects with the mesonic one where $\mu = u_c - u_T$ and $n=0$, and with the HTQ solution where 
$u_c=u_T$ at some nonzero $n$. As a consequence, the right panel of the figure shows a first order phase transition from the mesonic to the LTQ phase followed by a transition to the HTQ phase. From that figure it is difficult to say how smooth the LTQ-HTQ transition is.  Our results strongly suggest that the density, i.e., the first derivative of the free energy, is continuous (see also Fig.\ \ref{fig:NandOMvsMU}). Therefore, it is not a first-order transition. We shall later come back to the question whether it is a higher-order transition or a smooth crossover. 

In Fig.\ \ref{fig:chiral} we plot the chiral condensate (\ref{cond}) (using 
Eq.\ (\ref{NG0})) and the constituent quark mass (\ref{Mq}) as a function of $\tilde{\mu}$ for two different temperatures and four different values of the mass parameter. The values of the mass parameter are the same as used later for the phase diagrams in Sec.\ \ref{sec:Phases}. We see that at zero temperature there is a transition from the mesonic to the LTQ phase. This transition is of first order for small quark masses and becomes continuous for larger quark masses. The HTQ phase does not exist at zero temperature. At nonzero temperatures, here $\tilde{t}=0.1$, the
HTQ phase starts playing a role, and we find a mesonic-HTQ first order transition for small quark masses and the sequence mesonic-LTQ-HTQ, as already seen in Fig.\ \ref{fig:OMvsT} as a function of $\tilde{t}$. The curves shown here do not exhaust all possible scenarios. In particular, for larger temperatures there is a first-order LTQ-HTQ transition. This will become clear in 
Fig.\ \ref{fig:NandOMvsMU} and in the phase diagrams of the next subsection. In Fig.\ \ref{fig:chiral}
we also see that both chiral condensate and constituent quark mass (and in fact all thermodynamic quantities) approach the same value at large chemical
potential, independent of the value of the quark mass. This is expected 
of course since for sufficiently large chemical potentials (and also for large temperatures) the quark mass should be negligible. Geometrically speaking, this means that the flavor branes in the HTQ phase become more and more straight and 
approach the chirally symmetric embedding. In the special case $\tilde{t}=0$ it is the LTQ phase that approaches the straight-brane embedding: as $\tilde{\mu}$ 
becomes larger the connected branes approach the horizon while assuming a rectangular shape. These asymptotic limits are a good check for our numerical calculation. All these observations could also have been made by plotting different thermodynamic quantities, for instance the density.

Let us now comment on the quantities themselves that are plotted in Fig.\ \ref{fig:chiral}. Chiral condensate and constituent quark mass behave qualitatively similar, but are not connected by a simple relation. In the 
simplest version of the NJL model, they are proportional to each other, connected by a (dimensionful) coupling constant. In QCD, the constituent quark mass does not only originate from the chiral condensate since gluons play an important role as well. It is thus not unphysical to observe different behaviors of the 
two quantities. However, we have already argued that our "chiral condensate" is 
really constructed from a non-local operator and its interpretation is thus not obvious. Indeed, we observe for instance that asymptotically, where one might expect chiral symmetry to be effectively restored, we obtain a nonzero value. This value is indicated by a horizontal dashed line and is given by 
\be \label{qqbarasym}
\frac{\<\bar{q}q\>}{\<\bar{q}q\>_0}\to 
\exp\left[\frac{\lambda_0}{\ell}\left(\ell^2 t^2 \frac{8\pi^{5/2}}{3}\frac{\Gamma[5/3]}{\Gamma[1/6]}-\pi \tan\frac{\pi}{16}\right)\right] \, ,
\ee
where the arrow stands for the limit $\mu\to\infty$ or $t\to \infty$, and where we have used Eqs.\ (\ref{tuT}), (\ref{K1}), and (\ref{Adef2}). At fixed $\lambda_0$ and $\ell$, this expression grows exponentially with $t$. This 
$t$-dependence is left since we have not renormalized the Nambu-Goto part of the worldsheet action by the result of the straight-brane solution but with a vacuum term that does not depend on temperature, see discussion around Eq.\ (\ref{SNGstraight}). Interestingly, however, if $\ell$ is sent to zero at fixed $\lambda_0$ and $t$, the asymptotic chiral condensate, compared to the vacuum value, approaches zero, as one would expect. 
(Eq.\ (\ref{qqbarasym}) is only valid for the LTQ and HTQ phases, i.e., the mesonic piece of the curves in Fig.\ \ref{fig:chiral} does not at the same time go to zero.) 
The difficulties 
in the interpretation of the chiral condensate and the constituent 
quark mass in the Witten-Sakai-Sugimoto model are well known \cite{Aharony:2008an,McNees:2008km,Hashimoto:2008sr,Argyres:2008sw,Evans:2007jr}, and Fig.\ \ref{fig:chiral} underlines them. Nevertheless, 
one of our main points will be that the phase structure obtained in the present 
approach is very similar to the D3-D7 approach, where there {\it is} a straightforward concept of a chiral condensate and where the chiral condensate {\it is} zero in the chirally symmetric phase. This will be particularly obvious from the phase diagrams in the next subsection.

In Fig.\ \ref{fig:NandOMvsMU} we choose a fixed value of the mass parameter and plot the quark number density as a function of the chemical potential for various different temperatures in the left panel, and show the behavior of the (logarithm of the) pressure $P=-\Omega$ for one selected temperature in the right panel. The density curves 
can be understood as follows: at zero temperature (rightmost curve) the LTQ phase exists for arbitrarily large chemical potentials. As we increase the temperature, a point where the LTQ phase connects with the HTQ phase (red and blue curves in the figure), appears at large $\tilde{\mu}$ and moves all the way down to $\tilde{\mu}=0$. In doing so, it bends and drags the LTQ curve backwards. Once the origin is reached, the LTQ curve detaches from the HTQ phase and connects at two points to the mesonic solution ($n=0$), rather than interpolating between the 
mesonic and HTQ phases. For even larger temperatures (not shown here) this branch disappears and the LTQ solution ceases to exist. In the right panel we demonstrate that for certain temperatures there is a first-order LTQ-HTQ transition. For a better distinguishability of the curves, we have  normalized all free energies by the free energy of the HTQ phase. We clearly see that there is a continuous transition from the mesonic phase to the LTQ phase, followed by a first-order transition to the HTQ phase. 

\subsection{Speed of sound}
\label{sec:speed}

We have shown that there can be a continuous transition between the LTQ and HTQ phases. This continuity can be understood from the geometry in the bulk, but also manifests itself in the thermodynamic quantities considered so far. In particular, the quark number density is (within the very small numerical uncertainties) 
continuous across the transition, i.e., for all we know so far, the transition 
could be a smooth crossover. To analyze the smoothness of the transition, let us therefore compute second derivatives of the free energy. The speed of sound combines all possible second derivatives with respect to temperature and chemical potential and thus is a suitable quantity for our purpose. Comparable results of the speed of sound at nonzero temperature and chemical potential have been obtained in holography \cite{Gursoy:2017wzz} as well as in NJL models \cite{Bhattacharyya:2012rp,Marty:2013ita}. The speed of sound $c_s$ is defined as
\begin{equation}
    c_s^2 = \frac{\partial P}{\partial \epsilon} = \frac{n^2 \frac{\der s}{\der t} + s^2 \frac{\der n}{\der \mu} - n s \left(\frac{\der n}{\der t} + \frac{\der s}{\der \mu}\right)}{w 
    \left(\frac{\der n}{\der \mu}\frac{\der s}{\der t} - \frac{\der n}{\der t}\frac{\der s}{\der \mu}\right)}  \, ,  
    \label{csdef}
\end{equation}
where $s$ is the entropy density and $w=\mu n+st$ the enthalpy density, all partial derivatives with respect to $t$ are taken at fixed $\mu$ and vice versa, and
the derivative of the pressure $P=-\Omega$ with respect to the 
energy density $\epsilon = \Omega+\mu n + st$ is taken at fixed entropy per particle $s/n$. In the definition (\ref{csdef}) we have already replaced all thermodynamic quantities by their dimensionless counterparts. It is easy to check that this does not change the result for $c_s$ since all dimensionful constants and numerical prefactors cancel. We have computed the speed of sound using both expressions given in Eq.\ (\ref{csdef}) and found that the numerically cleanest result is obtained by using the right-hand side with all second derivatives evaluated as far as possible
in a semi-analytical way. These semi-analytical  expressions (analytical up to a numerical integration) are derived in appendix \ref{app1}.

\begin{figure}[t]
        
        \includegraphics[width=0.49\textwidth]{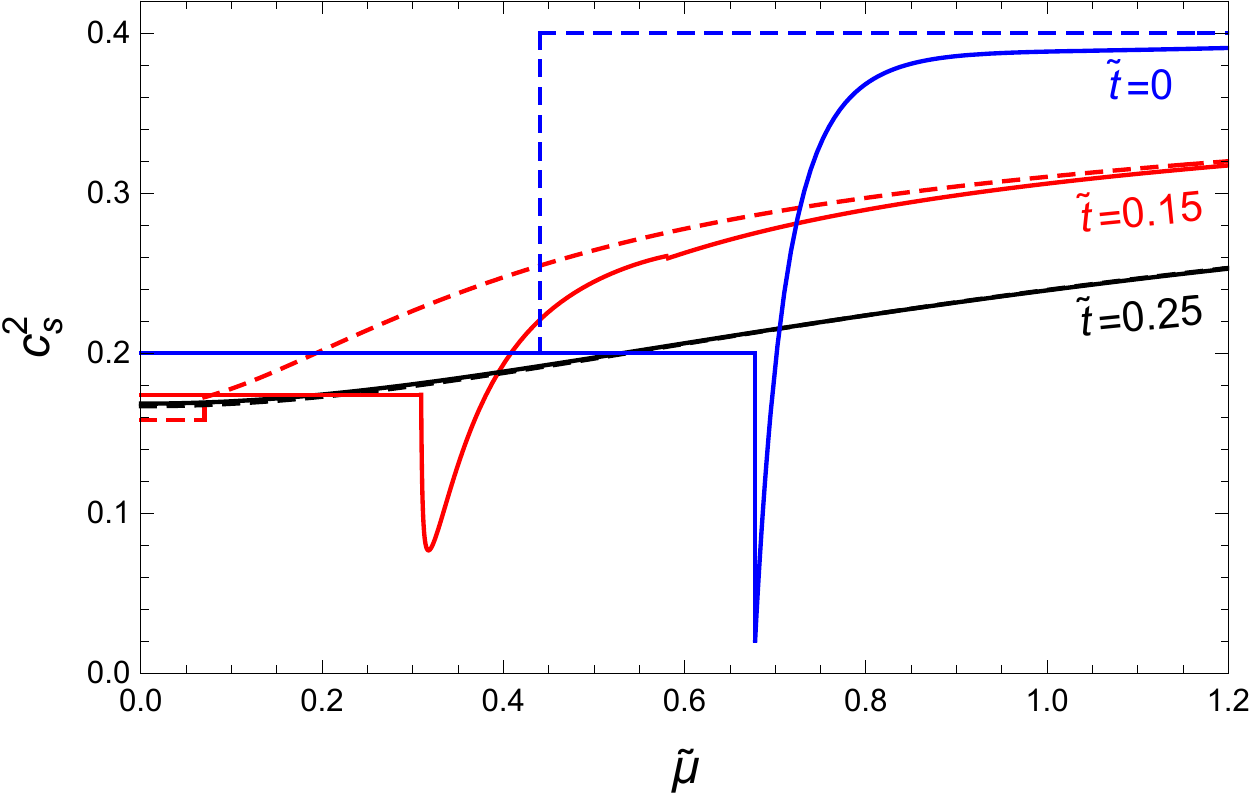}
        \includegraphics[width=0.49\textwidth]{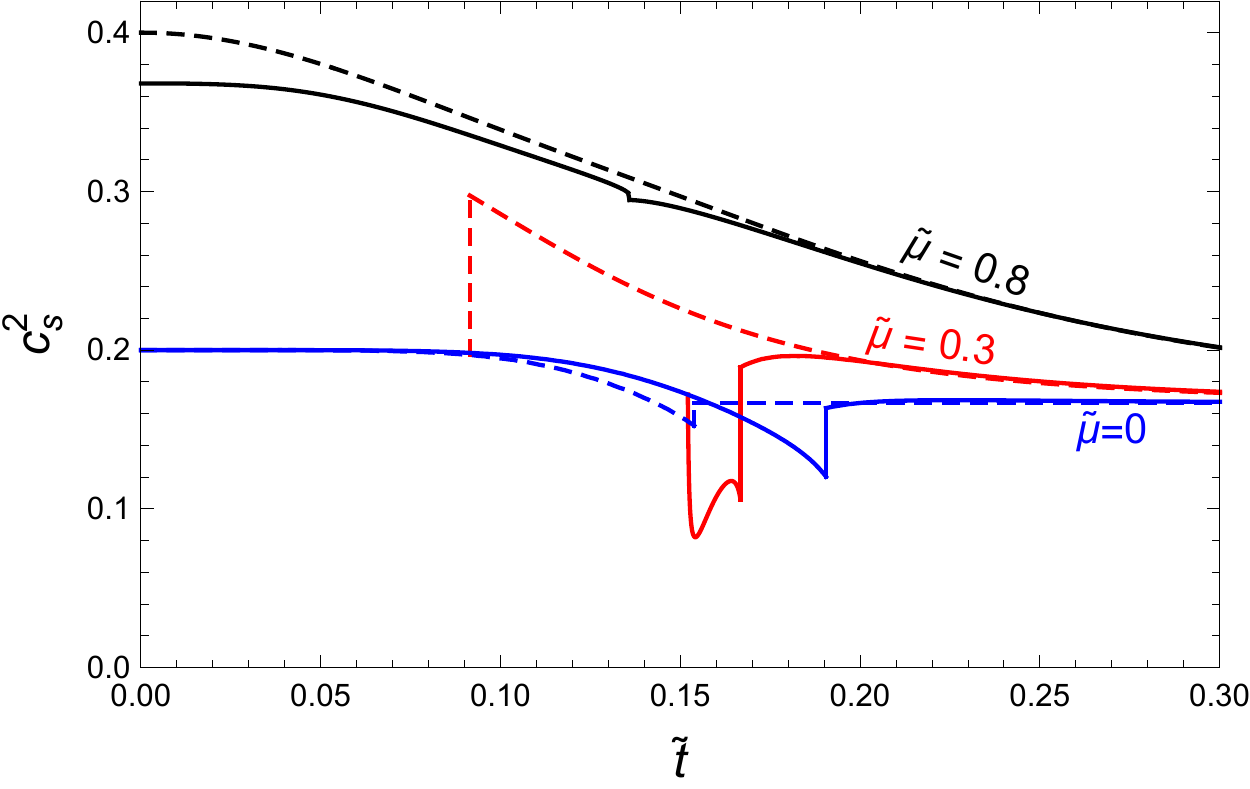}
        
     \caption{{\small Speed of sound (squared) in units of the speed of light as a function of $\tilde{\mu}$ at three different temperatures (left panel) and as a function of $\tilde{t}$ at three different chemical potentials (right panel) for a mass parameter $\tilde{\alpha}=0.04$. The dashed lines 
     are the corresponding results in the chiral limit $\tilde{\alpha}=0$. Here only the stable branches of the solutions are shown and discontinuities correspond to phase transitions.}}
     \label{fig:sound}
\end{figure}

The numerical results are shown in Fig.\  \ref{fig:sound}. In contrast to the plots in the previous subsection, here we only show the results for the stable phases, i.e., metastable and unstable branches are omitted. Each curve thus contains 
contributions from two or three of the mesonic, LTQ, and HTQ phases, separated by phase transitions. We see that for large temperatures or chemical potentials the speed of sound approaches the massless result, as expected. This is a further check of our results since the speed of sound of the chirally symmetric phase in the massless limit can be computed from a completely different, much simpler expression \cite{BitaghsirFadafan:2018uzs},
\be
\mbox{straight branes:} \quad c_s^2 = \frac{2}{5}\frac{u_T\sqrt{n^2+u_T^5}(n^2+5u_T^5)+\mu n(n^2+6u_T^5)}{(n^2+6u_T^5)(\mu n+2u_T\sqrt{n^2+u_T^5})} \, , 
\ee
where $n$ and $\mu$ are related by 
\be 
\mu=\frac{n^{2/5}\Gamma[3/10]\Gamma[6/5]}{\sqrt{\pi}}
- u_T\,{}_2 F_1\left[\frac{1}{5},\frac{1}{2},\frac{6}{5},-\frac{u_T^5}{n^2}\right] \, .
\ee
In particular, all curves approach $c_s^2=2/5$ for large $\tilde{\mu}$ at fixed $\tilde{t}$ and $c_s^2=1/6$ for large $\tilde{t}$ and fixed $\tilde{\mu}$. Due 
to the inherent scale $M_{\rm KK}$ and the lack of asymptotic freedom in this 
version of the model, these asymptotic values are different from the 
scale-invariant value $c_s^2=1/3$ that is assumed asymptotically in QCD.

Let us comment on the various phase transitions visible in Fig.\ \ref{fig:sound}. The $\tilde{t}=0$ curve in the left panel includes a mesonic-HTQ first-order transition which manifests itself in a discontinuity of the speed of sound. 
The $\tilde{t}=0.15$ curve shows a similar feature, but in this case it is a 
mesonic-LTQ transition with a smaller discontinuity. Since the numerics become challenging in the LTQ phase at very small $n$, it is difficult to compute the 
values of the speed of sound along the almost vertical segment of the $\tilde{t}=0.15$ curve, i.e., the size of the jump is difficult to predict.  
More importantly, the same curve also shows a small, barely visible, discontinuity
as we move from the LTQ to the HTQ geometry. In the vicinity of this transition, the numerics of the HTQ phase is completely under control. The reason is that, 
as mentioned above, the HTQ solution formally continues beyond the point $x_4(u_T)=0$ to unphysical, negative values of $x_4(u_T)$. Therefore, the point 
$x_4(u_T)=0$ is mathematically not very special. The numerics of the LTQ phase is somewhat more difficult as one approaches the point $u_c\to u_T$, but the results for the speed of sound suggest no particular problems either. Therefore, we can be confident that this small discontinuity is not an artifact of the numerics. As a result, 
the LTQ-HTQ transition shown here is not a smooth crossover but a second-order phase transition, although with a very small discontinuity in the 
second derivatives of the thermodynamic potential. The same feature is visible in the $\tilde{\mu}=0.8$ curve in the right panel. The $\tilde{\mu}=0.3$ curve in this panel shows the same mesonic-LTQ transition as in the left panel, but now  followed by a first-order LTQ-HTQ transition, as discussed in the context of 
Fig.\ \ref{fig:NandOMvsMU}.

\subsection{Phase diagrams}
\label{sec:Phases}

\begin{figure}[t]
    \centering{        
       \includegraphics[width=0.49\textwidth]{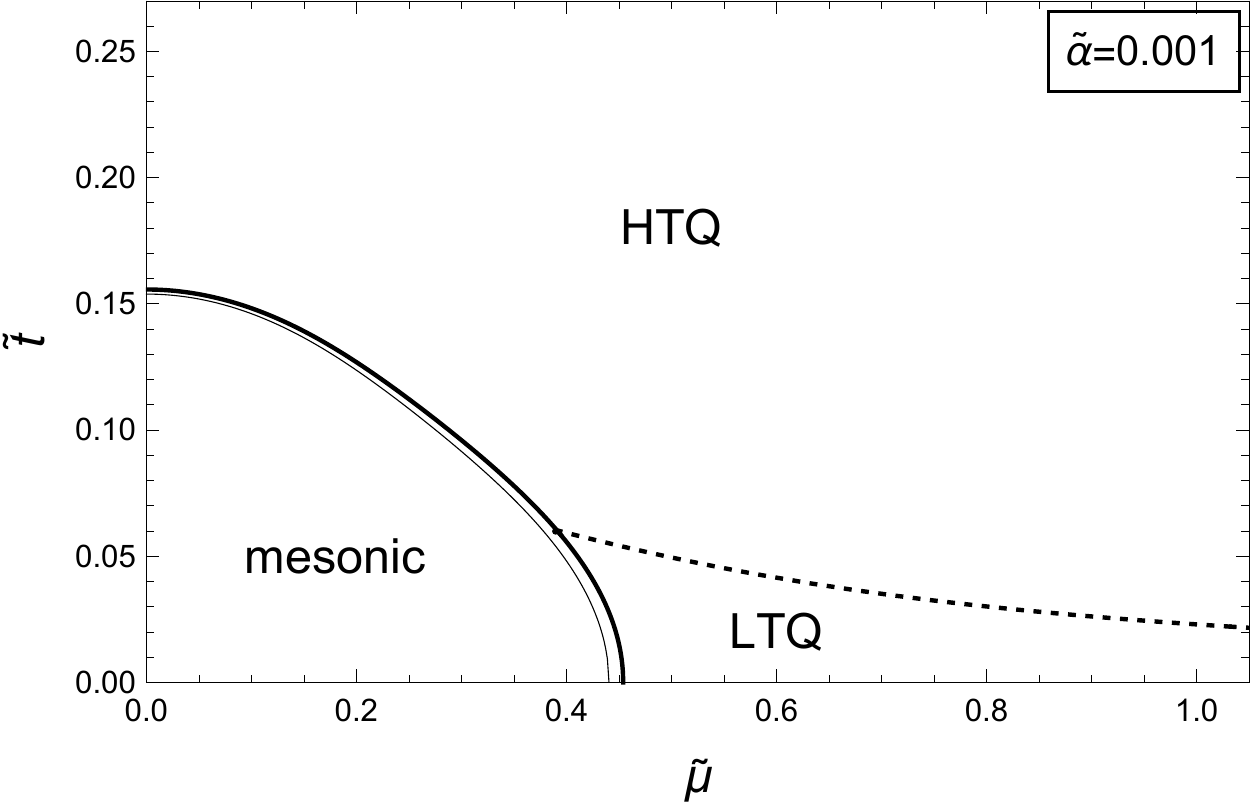}
        \includegraphics[width=0.49\textwidth]{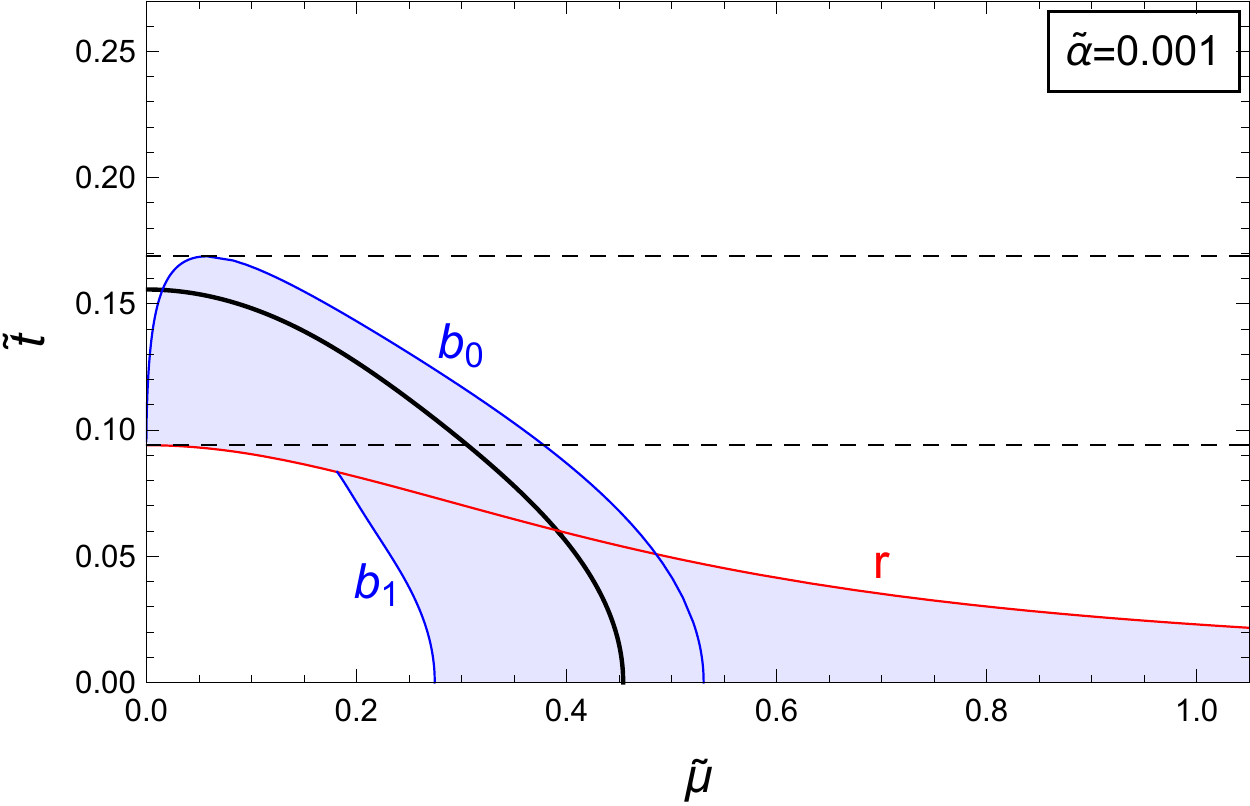} 
        
       \includegraphics[width=0.49\textwidth]{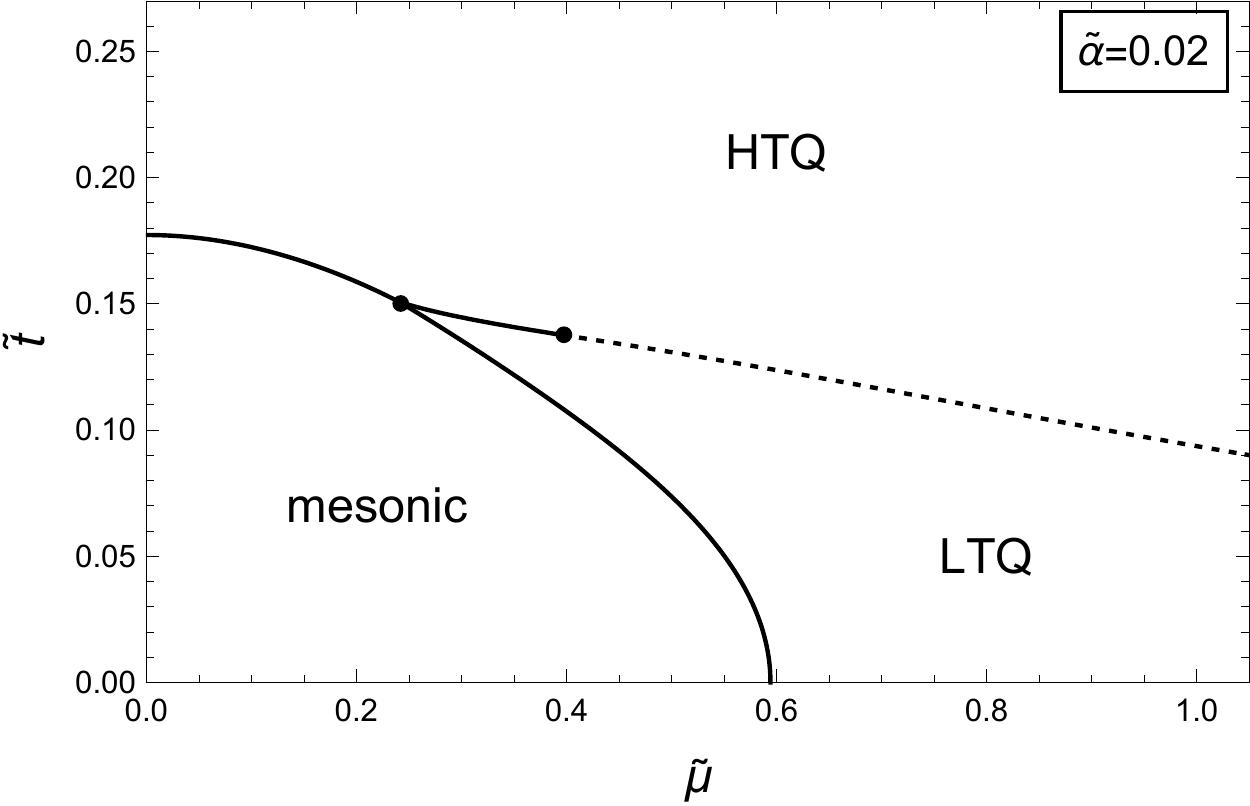}
        \includegraphics[width=0.49\textwidth]{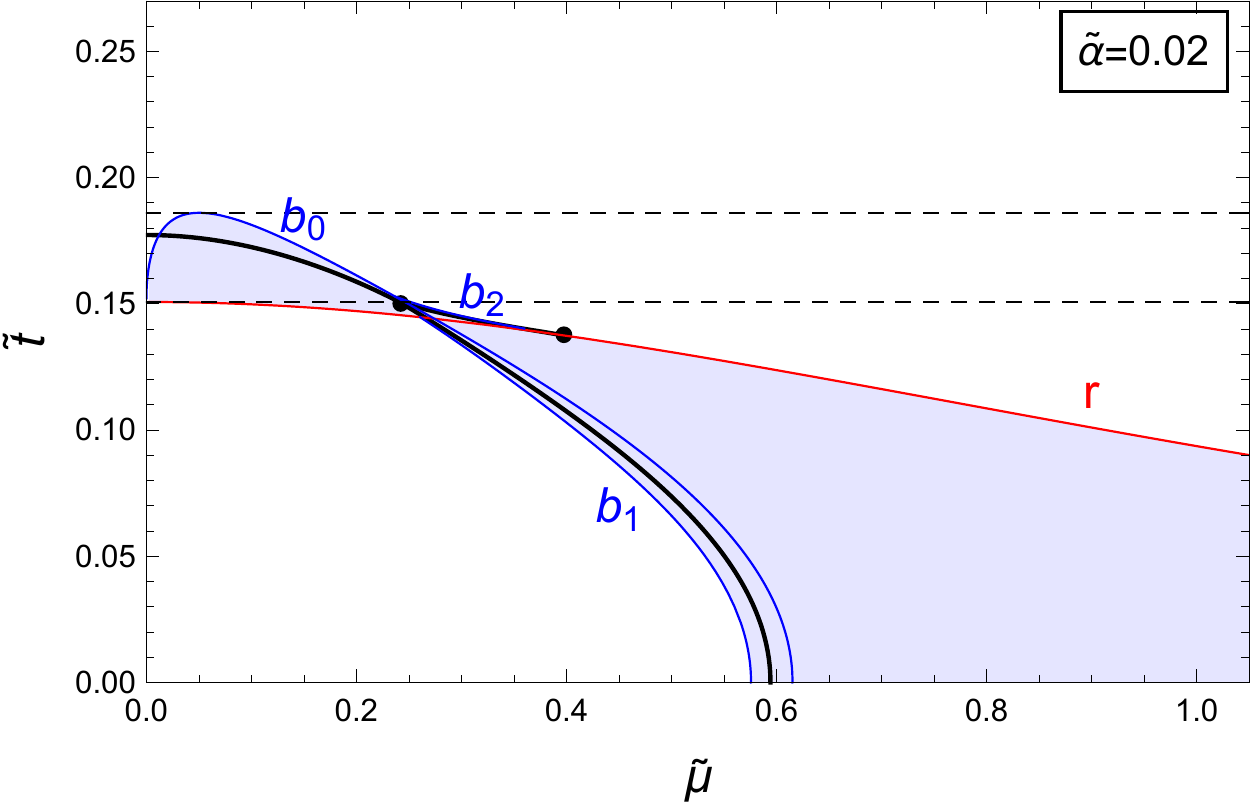}
        
       \includegraphics[width=0.49\textwidth]{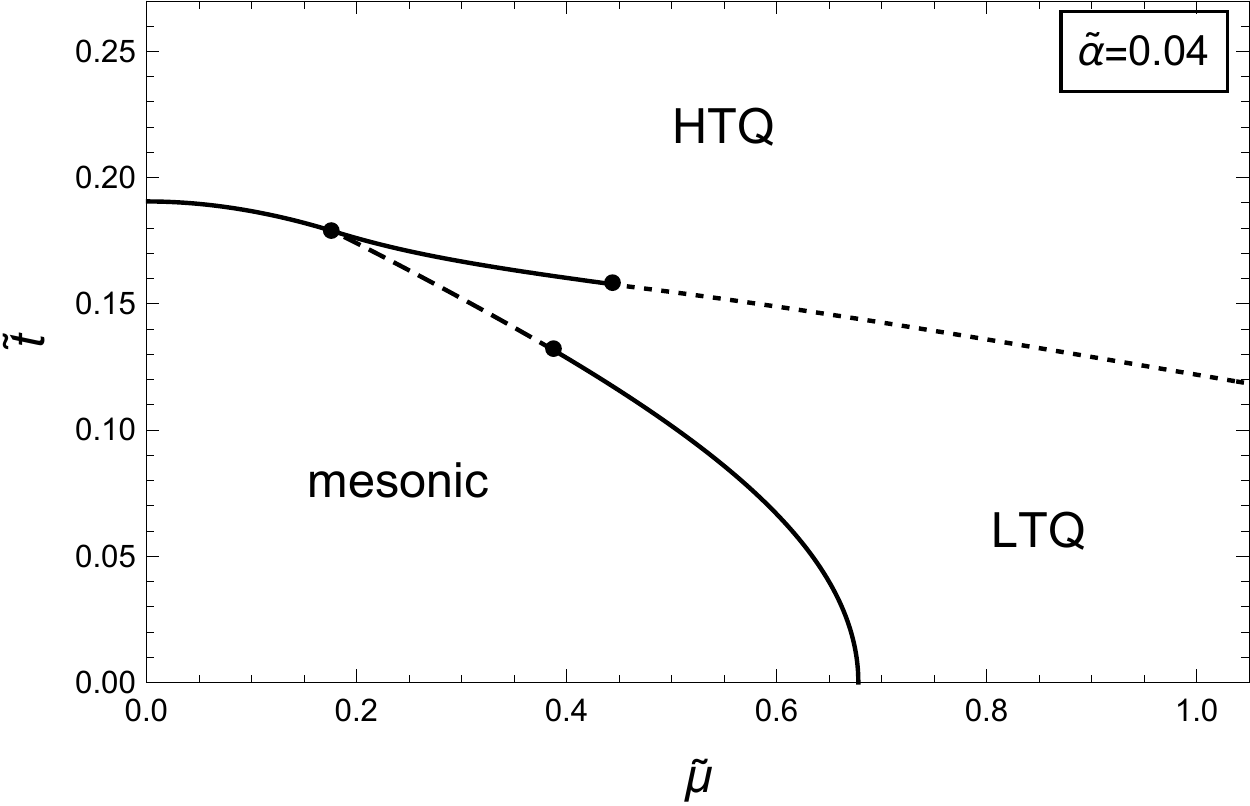}
        \includegraphics[width=0.49\textwidth]{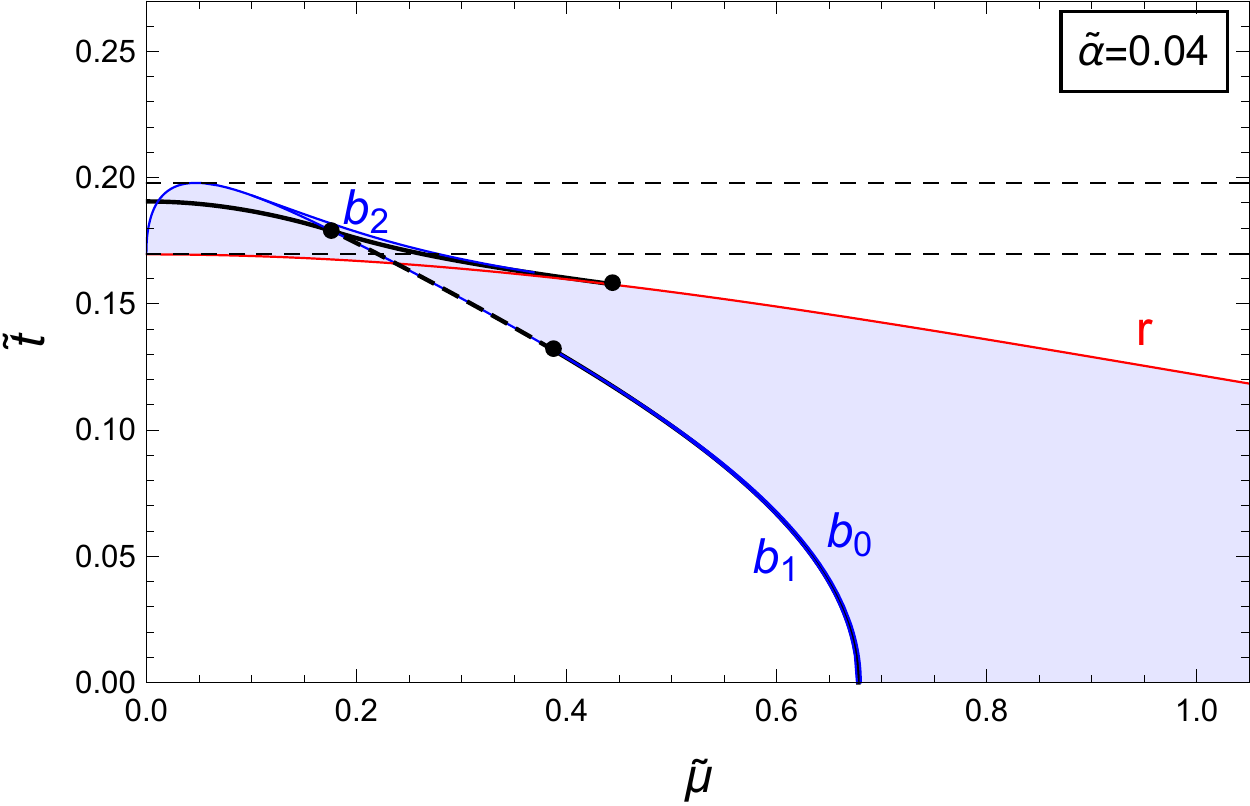}
        
       \includegraphics[width=0.49\textwidth]{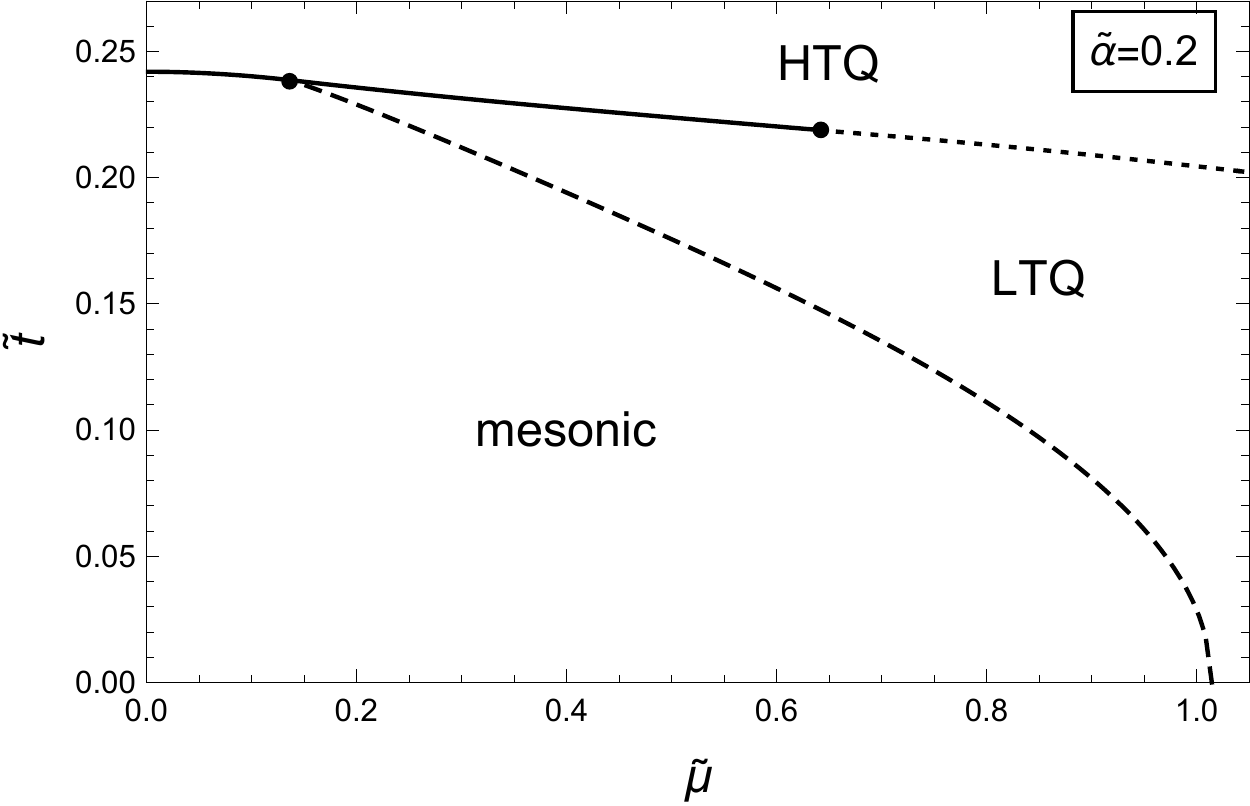}
        \includegraphics[width=0.49\textwidth]{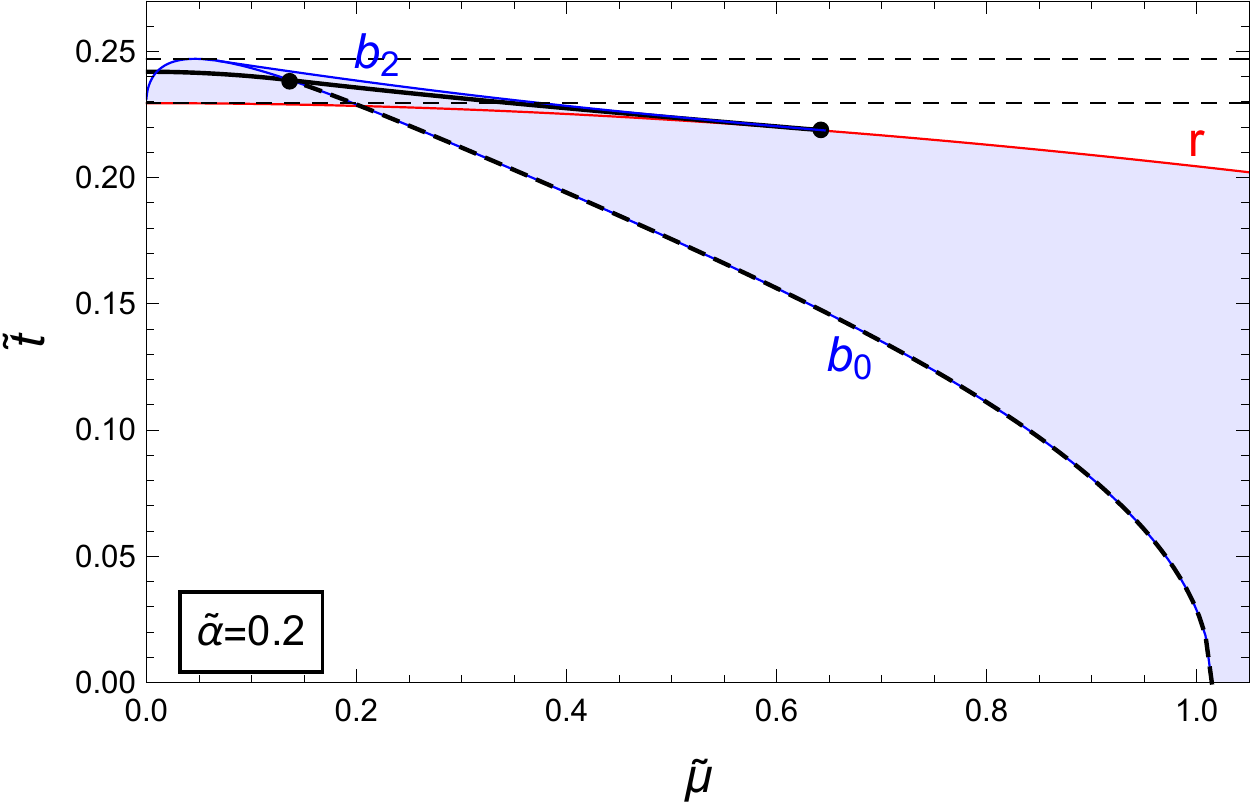}
    }
     \caption{{\small Left panels: Phase diagrams in the plane of dimensionless temperature $\tilde{t}$ and quark chemical potential $\tilde{\mu}$ for increasing quark mass parameter $\tilde{\alpha}$ from top to bottom and fixed coupling $\tilde{\lambda}=15$. Solid lines are discontinuous phase transitions, while transitions across dashed and dotted lines are continuous. The thin solid line in the upper left panel is the chiral phase transition in the massless limit. 
     Right panels: Same as left, but including curves that indicate regions of metastable and unstable solutions. 
     For detailed explanations see text. The shaded region shows where the LTQ configuration exists.
     }}
     \label{fig:Phasediagrams}
     
     \vspace{-1.5cm}
\end{figure}

The main result of this paper is shown in Fig.\ \ref{fig:Phasediagrams}.
This figure shows four different phase diagrams, which summarize the various phase transitions discussed in the previous subsections. Before we proceed to the discussion of the results, let us briefly explain how we have calculated the phase transition lines. It would obviously be very tedious to determine the phase transitions point by point by calculating 
the free energies for all possible phases as functions of $\tilde{\mu}$ or $\tilde{t}$. Moreover, it would then be easy to miss certain features of the phase structure, unless the free energies are computed on a very fine grid in $\tilde{\mu}$ and $\tilde{t}$, which would be very laborious. It is thus much more advantageous to compute the phase transition lines in a more direct way. For instance, for a first-order phase transition curve we have solved a system of equations simultaneously that combines the equations of the two respective phases that coexist at this phase transition line plus the condition that they have the same free energy. All curves in Fig.\ \ref{fig:Phasediagrams} have been obtained in this direct way, i.e., by suitably combining the stationarity equations and various conditions for the free energies. The figure contains four phase diagrams for four different values of the (rescaled) mass parameter $\tilde{\alpha}$, with the quark mass increasing from top to bottom rows. As in the previous section, $\tilde{\lambda}=15$. The four phase diagrams in the left panels are duplicated in the right panels, but with additional auxiliary curves that are helpful for the understanding (and for reproducing) the phase structure since they indicate the regions of unstable and metastable solutions. In particular, these curves define the regions where each of the three phases exist and keep track of the rich 
behavior of the LTQ solution:
\begin{itemize}
\item The mesonic phase exists everywhere below the upper horizontal dashed line.
Between the two horizontal dashed lines it is two-valued, while below the lower dashed line it has a single solution, see also Fig.\ \ref{fig:OMvsT}. All its equilibrium properties are independent of the chemical potential since the quark density is zero. 

\item The LTQ phase exists in the shaded area. It connects to the HTQ phase on the (red) curve labeled by $r$, on which $u_c=u_T$ and which asymptotes to the horizontal axis for $\tilde{\mu}\to \infty$. It connects to the mesonic phase on the (blue) curve labeled by $b_0$, where $n=0$. In all four diagrams, $b_0$ 
is a continuous line with endpoints at both axes. At the (blue) curves labeled by $b_1$ and $b_2$, the solution "turns around", i.e., along these curves
$\frac{\partial n}{\partial \mu}=\infty$, see also Fig.\ \ref{fig:NandOMvsMU}. 

\item The HTQ phase exists everywhere above the (red) curve labeled by $r$. On this curve, $x_4(u_T)=0$ and it connects to the LTQ phase. This solution is always single-valued (provided only the physical solutions with $x_4(u_T)\ge 0$ are considered).
\end{itemize}

The four values of the mass parameter are chosen to represent prototypical examples of the four different topologies of the phase diagram we have found. For small current quark masses the phase diagram is qualitatively the same as in the chiral limit: there is a discontinuous transition between the mesonic phase, which has zero quark number, $n=0$, to the phases with nonzero quark number $n>0$. In contrast to the chiral limit, the $n>0$ region now contains the LTQ phase, whose area grows as we increase the current quark mass. We have depicted the transition between LTQ and HTQ phases by a dotted line. This indicates that the transition is continuous, i.e., density and entropy do not jump when we cross from one phase to the other. However, we have seen in the calculation of the speed of sound that second derivatives have a (very small) discontinuity. This discontinuity might be a consequence of constructing the LTQ phase with the help of string sources, which possibly is an approximation to a phase with a 
more complicated flavor brane embedding, as already discussed at the end of Sec.\ \ref{sec:LTQ}. We cannot exclude that a more complete treatment removes this second-order discontinuity and results in a smooth crossover. 

\begin{figure}[t]
    \centering
    \includegraphics[width=0.5\textwidth]{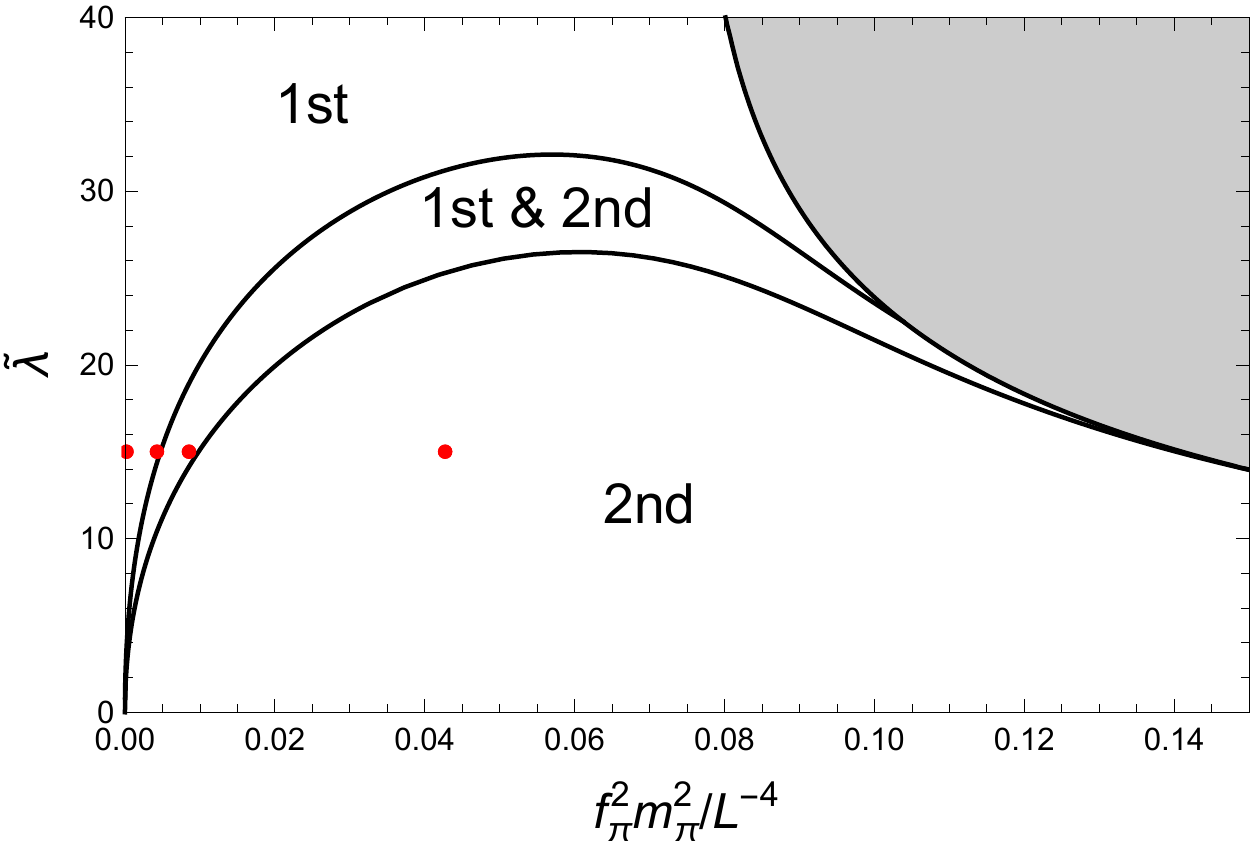}
    \caption{{\small Different topologies of the phase diagrams in the plane of rescaled 't Hooft coupling $\tilde{\lambda}$ and quark mass parameter, classified according to the mesonic-LTQ transition, which can be discontinuous everywhere ("1st"), discontinuous at small temperatures and continuous at large temperatures ("1st \& 2nd") or continuous everywhere ("2nd"). The shaded area is excluded since our approximation breaks down. The 4 (red) dots mark the parameter pairs chosen for the phase diagrams in Fig.\ \ref{fig:Phasediagrams}. The translation from the mass parameter $\tilde{\alpha}$ to the more physical 
    quantity $f_\pi^2 m_\pi^2$ is obtained with the help of Table \ref{table1}. }} 
    \label{fig:lambdabeta}
\end{figure}

As we move to larger values of the current quark mass, a first-order phase transition line develops at intermediate temperatures. The endpoint of this line is somewhat difficult to calculate because the free energies of the phases are extremely close to each other in the vicinity of this point. Therefore, it is also difficult to calculate the critical value of the mass parameter where this first-order line starts to appear. Our numerics suggest $\tilde{\alpha}\simeq 0.007$. Increasing the mass parameter further, the
original first-order chiral transition line breaks up, and a continuous transition between the mesonic and the LTQ phases appears. The exact location of the resulting endpoint of the first-order mesonic-LTQ transition is again prone to some numerical uncertainties. Eventually, at even larger mass, the  mesonic-LTQ transition becomes continuous throughout. 

The "evolution" of the phase diagram discussed so far is obtained at fixed (rescaled) 't Hooft coupling $\tilde{\lambda}$, and one may ask how the phase structure changes if also $\tilde{\lambda}$ is varied. (Keeping in mind that our classical gravity approximation is only valid at strong 't Hooft coupling $\lambda$ and results for small values of the coupling are an extrapolation.) The 
result is shown in Fig.\ \ref{fig:lambdabeta}, which indicates the different topologies of the phase diagram in the $\tilde{\lambda}$-$\tilde{\alpha}$-plane. 
Recall that since we were able to absorb the other model parameters $\ell$ and $M_{\rm KK}$ in our rescaled variables, $\tilde{\lambda}$ and $\tilde{\alpha}$ are the only independent parameters, and thus Fig.\ \ref{fig:lambdabeta} presents a complete systematic study of the phase structure. This figure contains an unphysical region for large values of the mass parameter, shown as a shaded area. As we enter this area, a region in the $\tilde{t}$-$\tilde{\mu}$ plane appears where there is no stable phase in our approximation. The reason is that the  $\cup$-shaped solution ceases to exist for small temperatures (for all temperatures as we move deeper into the shaded area), as already pointed out in Sec.\ \ref{sec:thermo}. This shortcoming of our approach is no surprise since we have only added a term linear in the current quark mass to the action. 

For the horizontal axis in this figure we have translated the mass parameter $\tilde{\alpha}$ into the more physical quantity $f_\pi^2m_\pi^2/L^{-4}$ via Table \ref{table1} with $N_c=3$. 
For a rough estimate, let us use $L\simeq 10^{-3}\, {\rm MeV}^{-1}$, which was 
obtained from a fit to nuclear matter properties at saturation density \cite{BitaghsirFadafan:2018uzs}. Then, with the actual QCD vacuum values $f_\pi\simeq 93 \, {\rm MeV}$ and $m_\pi\simeq 140\, {\rm MeV}$ we have $f_\pi^2m_\pi^2/L^{-4} \simeq 1.7\times 10^{-4}$. At $\tilde{\lambda}=15$ this corresponds to $\tilde{\alpha}\simeq 8\times 10^{-4}$, slightly smaller than even the smallest mass parameter used in Fig.\ \ref{fig:Phasediagrams}. Thus all interesting features of our phase diagram occur in the heavy QCD regime. We may also use this fit to get a rough idea of the scales in the phase diagrams. With the help 
of Table \ref{table1} we find that for the chiral limit the $\mu=0$ phase transition is at about $T_c\simeq 150\, {\rm MeV}$, while the $t=0$ (discontinuous) phase transition occurs at a quark chemical potential of about $\mu_c\simeq 520\, {\rm MeV}$. 
The heaviest quark mass parameter considered in Fig. \ref{fig:Phasediagrams}
corresponds to $f_\pi^2m_\pi^2$ being about 250 times its QCD value. In this case the corresponding critical values are $T_c\simeq 240\, {\rm MeV}$ and (now for a continuous transition) $\mu_c\simeq 1200\, {\rm MeV}$. 

It is tempting to compare our results to lattice QCD, where at large quark masses and large coupling the phase structure can be computed even for nonzero chemical potentials \cite{Fromm:2011qi,Philipsen:2019qqm}. If we take our heaviest case and ignore the fact that our continuous phase transitions have higher-order discontinuities, the phase diagram has a intriguing similarity to the heavy QCD diagram, see for instance Fig.\ 1 of Ref.\ \cite{Philipsen:2019qqm}. The similarity becomes even more striking when we go to our intermediate regime where 
there is a low-temperature discontinuous transition with a critical endpoint.  However, this comparison has to be taken with some care for at least two reasons. Firstly, our calculation does not include baryons. Therefore, our low-temperature mesonic-LTQ transition that ends in a critical point is different from 
the liquid-gas transition in Fig.\ 1 of Ref.\ \cite{Philipsen:2019qqm}. Secondly, heavy QCD becomes pure Yang-Mills in the limit of infinitely heavy quarks. In this limit, the first-order transition at large temperature is a strict deconfinement transition. In our approach, however, we have decoupled the gluon dynamics such that our high-temperature transition does not seem to know anything about confinement. Nevertheless, the resemblance of the phase diagrams with heavy QCD is striking and suggests further studies for a better understanding of the parallels and differences to QCD, most notably by including baryons into our holographic 
calculation.  

A more concrete comparison can be made to the phase diagrams obtained in holographic studies with a D3-D7 setup. We have already pointed out the very close analogy of our three flavor brane configurations to the corresponding D3-D7 geometries. This analogy is also borne out in the phase diagrams, as one can see for instance by comparing our results to Fig.\ 5 in Ref.\ \cite{Evans:2010iy} (where a nonzero magnetic field was introduced in order to break conformal symmetry). This
particular figure is obtained with zero current quark mass, but in that model 
the quark mass does not seem to change the topology of the phase diagram except for turning the chiral phase transition into a smooth crossover. The first-order transition, including its critical endpoint, as well as the overall structure is qualitatively the same as in our heaviest case, see lowest panels in Fig.\ \ref{fig:Phasediagrams}, if the meson-melted D3-D7 phase is identified with our LTQ phase. This very close resemblance is somewhat surprising since, as already discussed, in the present model the identification of an order parameter for chiral symmetry breaking is much less straightforward than in the D3-D7 approach. Turning this argument around, we can take the result as a reassurance for our approach, despite the open conceptual questions regarding the implementation of the chiral condensate in the Witten-Sakai-Sugimoto model.

\section{Summary and outlook}
\label{sec:Outlook}

We have included a nonzero current quark mass in the study of the phase diagram in holographic QCD. Working in the deconfined geometry and the decompactified limit of the Witten-Sakai-Sugimoto model, we have added a mass correction to the action analogous to chiral perturbation theory. To this end, we have made use of earlier works that identified the chiral condensate with the expectation value
of a non-local, gauge-invariant open Wilson line operator. The holographic dual of this expectation value corresponds to a worldsheet instanton, and we have 
included this contribution for the first time in a fully consistent way, calculating its effect on the embedding of the flavor branes in the background geometry. We have identified three different embeddings. Most notably, we have pointed out that the current quark mass stabilizes a brane configuration with string sources, which previously had only  been discussed in the chiral limit, where it is never stable.
Evaluating the classical equations of motion and the stationarity conditions of the on-shell action, we have systematically studied the equilibrium phase structure in the plane of temperature and baryon chemical potential. The effect of heavy quarks is to break up the chiral phase transition line  and introduce a critical endpoint at high temperatures. At intermediate values of the quark mass there is an additional endpoint at lower temperatures. This phase structure is qualitatively similar to heavy QCD on the lattice. We have also pointed out that the phase structure as well as the geometric structure of the different embeddings is strikingly similar to a different holographic approach based on the D3-D7 setup. This similarity is much less obvious in the chiral limit. 

The main significance of our work is to make the study of equilibrium phases 
in the Witten-Sakai-Sugimoto model more realistic. Therefore, our results can be used as the basis for further studies of strongly-coupled hot and dense matter. The most obvious extension is to include baryonic matter, possibly as a first step from simple approximations based on pointlike baryons \cite{Bergman:2007wp}, or from more sophisticated approximations using nonzero-width, interacting instantons \cite{BitaghsirFadafan:2018uzs}. This  extension would be very interesting for the phase structure itself but also for more specific questions such as the quark-hadron continuity at low temperatures \cite{BitaghsirFadafan:2018uzs}
and for a realistic equation of state for dense matter, to be applied to the 
physics of neutron stars. 

Our work also suggests several other directions for future research. For instance, we have only considered the case of a degenerate quark mass for all flavors, such that the number of flavors played no important role. The generalization to different quark masses is straightforward and could have non-trivial effects on the phase structure of the system. 
We have also pointed out and reinforced several conceptual problems related to the introduction of a current quark mass in the Witten-Sakai-Sugimoto model. This concerns for instance the identification of chiral condensate, constituent quark mass, and order parameter for chiral symmetry breaking. Our results for the phase diagram, which 
seem very sensible in comparison to other model approaches and even compared to actual QCD, suggest that one might simply go ahead with the present approach. Nevertheless, some conceptual progress would be desirable as well. It might be useful to repeat our study employing another approach that is 
suggested in the literature for the introduction of a mass term, based on a tachyonic scalar field. An improvement to our study on the conceptual level would also be to find a smooth flavor brane configuration that replaces our construction with string sources and a cusp in the embedding of the branes.

\acknowledgments

It is a pleasure to thank Kazem Bitaghsir Fadafan, Yuri Cavecchi, Nick Evans, Josef Leutgeb, Owe Philipsen, Anton Rebhan, Mart\'in Schvellinger, and Motoi Tachibana for interesting comments and valuable insights. This work is supported by the Leverhulme Trust under grant no RPG-2018-153. The work of N.K.\ has additionally been supported  by the National Agency for the Promotion of Science and Technology of Argentina (ANPCyT-FONCyT) Grants PICT-2017-1647 and PICT-2015-1525. A.S.\ is supported by the Science \& Technology Facilities Council (STFC) in the form of an Ernest Rutherford Fellowship.

\appendix
\section{Calculating the speed of sound}
\label{app1}

Here we explain some useful details regarding the calculation of the speed of sound, defined in \eqref{csdef}. As an input for this calculation we need the first derivatives of the thermodynamic potential,  $n$ and $s$. The number density $n$ appears directly in our calculation and needs no further comment, while for the entropy density $s$ we derive a semi-analytical expression in this appendix. Moreover, we need the second derivatives of the thermodynamic potential $\frac{\partial n}{\partial \mu}$, $\frac{\partial s}{\partial t}$, and the mixed derivative $\frac{\partial s}{\partial \mu}= \frac{\partial n}{\partial t}$. We shall derive a semi-analytical expression for the number susceptibility $\frac{\partial n}{\partial \mu}$. This is also useful for the phase diagrams in Fig.\ \ref{fig:Phasediagrams} since some of the curves in the right panels require 
a simultaneous solution of the stationarity equations and the equation $\frac{\partial \mu}{\partial n} = 0$. We refrain from deriving semi-analytical expressions for $\frac{\partial s}{\partial t}$ and $\frac{\partial n}{\partial t}$ because the expressions would be very lengthy. Since $n$ and $s$ are known, these derivatives can be computed purely numerically from finite differences without significant problems.  

The (dimensionless) entropy density is 
\begin{equation}
    s = - \frac{\der \Omega}{\der t} = - 2 \frac{u_T}{t}  \frac{\der \Omega}{\der u_T} \, , 
\end{equation}
where we have used Eq.\ (\ref{tuT}). 
With the DBI Lagrangian 
\be
{\cal L}_0 = u^{5/2} \sqrt{1 + u^3 f_T x_4'^2 - \hat{a}_0'^2}
\ee
we compute for the LTQ configuration
\bea
\frac{\der \Omega}{\der u_T} &=& \frac{\der}{\der u_T}\left[\int_{u_c}^{\infty} du \, {\cal L}_0  - \frac{A}{2\lambda_0} +n(u_c-u_T) -n\hat{a}_0(u_c)\right]  \non[2ex]
&=& \int_{u_c}^{\infty} du \left[ \frac{\der {\cal L}_0 }{\der x_4'}\frac{\der x_4'}{\der u_T}
+ \frac{\der{\cal L}_0 }{\der \hat{a}_0'}\frac{\der \hat{a}_0'}{\der u_T}
+ \frac{\der {\cal L}_0 }{\der u_T}
 - A \left(\phi_T \frac{\der x_4'}{\der u_T} + 
 x_4' \frac{\der \phi_T}{\der u_T}\right)\right] -n - n\frac{\der \hat{a}_0(u_c)}{\der u_T}\non[2ex]
&=&   \left(k \frac{\der x_4}{\der u_T}
- n \frac{\der \hat{a}_0}{\der u_T}\right)_{u=u_c}^{u=\infty}
+ \int_{u_c}^{\infty} du \left(\frac{\der {\cal L}_0 }{\der u_T}
 - A  x_4' \frac{\der \phi_T}{\der u_T}\right)-n - n\frac{\der \hat{a}_0(u_c)}{\der u_T} \, , 
\eea
where, in the last step, we have integrated by parts and taken into account the equations of motion. Note that the implicit dependence on $u_T$ through $u_c$ 
does not contribute since the derivative of $\Omega$ with respect to $u_c$ 
vanishes at the stationary point. Computing derivatives in this way, i.e., 
via ${\cal L}_0$, which is a functional of $x_4(u)$ and $\hat{a}_0(u)$, is equivalent to using the explicit form of the free energies in 
Sec.\ \ref{sec:Solutions}.
Since $x_4(u_c)$, $x_4(\infty)$ and $\hat{a}_0(\infty)$ are held fixed and thus their derivatives with respect to $u_T$ vanish, and using 
\begin{equation}
\frac{\der \phi_T}{\der u_T} = \frac{1}{u_T} \left[
\phi_T(u)-\frac{u}{\sqrt{f_T(u)}}\right]  \, ,   
\end{equation}
we find 
\begin{equation}
    s_{\Ydown} = \frac{2}{t} \left\{
    n u_T + \int_{u_c}^{\infty} du \, x_4' \left[
    \frac{3 u_T^3}{2 u^3} \frac{A \phi_T + k}{f_T} + A \left(
    \phi_T - \frac{u}{\sqrt{f_T}}
    \right)
    \right]
    \right\}.
\end{equation}
For small temperatures, since $u_T\propto t^2$, the entropy grows linearly with 
$t$. One easily finds that for the mesonic phase the entropy density $s_\cup$ has the same form with $n=0$. In this phase the speed of sound is simply given 
by 
\be
c_{s,\cup}^2 = \frac{s_\cup}{t}\left(\frac{\partial s_\cup}{\partial t}\right)^{-1} \, . 
\ee
For the HTQ phase the analogous calculation gives 
\begin{eqnarray}
    s_{\sqcup} &=& \frac{2}{t} \left\{ A \phi_T(u_T) x_4(u_T) +
    u_T^{7/2} \sqrt{1+\frac{n^2}{u_T^5} - \frac{4A^2}{9u_T^6}}
    \right. \nn \\
    &&\left. + \int_{u_T}^{\infty} du \, x_4' \left[
    \frac{3 u_T^3}{2 u^3} \frac{A \phi_T + k}{f_T} + A \left(
    \phi_T - \frac{u}{\sqrt{f_T}}
    \right)
    \right]
    \right\} \, .
\end{eqnarray}
Next we compute the number susceptibility. For the LTQ phase let us define 
\begin{equation}
    f_A(n,u_c,A) \equiv 
    A - \frac{2\alpha}{\lambda_0^2}\exp\left(2 \lambda_0
    \int_{u_c}^{\infty} du \,\phi_T x_4' \right) \, , \qquad   
    f_k(n,u_c,A) \equiv \frac{\ell}{2} -  \int_{u_c}^{\infty} du \, x_4' \, . 
\end{equation}
We compute $\frac{\partial \mu}{\partial n}$ at fixed $t$ along 
the surface given by the constraints $f_A=f_k=0$. These constraints are the first two relations in Eq.\ (\ref{Amudefstring}). 
To this end, we first write on account of the implicit function theorem
\begin{equation}
    \left(\frac{\der A}{\der n} , \frac{\der u_c}{\der n}\right) = 
    - \left(\frac{\der f_A}{\der n} , \frac{\der f_k}{\der n}\right)
    \left(\begin{array}{cc}
    \frac{\der f_A}{\der A} & \frac{\der f_k}{\der A} \\
    \frac{\der f_A}{\der u_c} & \frac{\der f_k}{\der u_c}
    \end{array} \right)^{-1}.
\end{equation}
The chemical potential $\mu$ is a function of $n$, $u_c$, and $A$ via the third relation in Eq.\ (\ref{Amudefstring}). Hence we can write 
\begin{equation}
    \frac{d \mu}{d n} = \frac{\der \mu}{\der n} + 
    \frac{\der \mu}{\der A}\frac{\der A}{\der n} + 
    \frac{\der \mu}{\der u_c}\frac{\der u_c}{\der n} \, . \label{dmudn}
\end{equation}
(Temporarily, we have employed the notation of a total derivative on the left-hand side, to avoid confusion with the explicit derivative on the right-hand side.) All partial derivatives can be taken semi-analytically by taking the 
derivatives of the integrands in the various integrals. The result is a complicated combination of numerical integrals which however is easily handled 
numerically. The number susceptibility $\frac{\partial n}{\partial \mu}$ is then simply obtained by the inverse of Eq.\ (\ref{dmudn}). 

It remains to compute the number susceptibility in the HTQ phase. This is somewhat simpler because here $u_c=u_T$ and thus there is one fewer dynamical variable. We
define 
\begin{equation}
    f_A(n,A) \equiv  A-\frac{2\al}{\lambda_0^2}
    \exp \left[ 2\lambda_0\left( \frac{\ell}{2} \phi_T(u_T) +  \int_{u_T}^\infty du\,  \left[\phi_T(u) - \phi_T(u_T) \right] x_4'(u)\right) \right] \, ,
\end{equation}
such that from the constraint $f_A=0$, see Eq.\ (\ref{Adef2}), we have 
\be
    \frac{\der A}{\der n} = - \frac{\der f_A}{\der n} 
    \left(\frac{\der f_A}{\der A}\right)^{-1} \, , 
\ee
which is used to compute  
\be
     \frac{d \mu}{d n} = \frac{\der \mu}{\der n} + 
    \frac{\der \mu}{\der A}\frac{\der A}{\der n} \, , 
\ee
where the explicit derivative of $\mu$ with respect to $n$ is obtained from the second relation in Eq.\ (\ref{Adef2}). 
This concludes the collection of all derivatives for all three phases needed for the calculation of the speed of sound according to the strategy laid out at the beginning of this appendix.


\bibliographystyle{JHEP}
\bibliography{refs}

\providecommand{\href}[2]{#2}\begingroup\raggedright\begin{thebibliography}{10}

\bibitem{Sakai:2004cn}
T.~Sakai and S.~Sugimoto, {\it {Low energy hadron physics in holographic QCD}},
   {\em Prog. Theor. Phys.} {\bf 113} (2005) 843--882,
  [\href{http://arxiv.org/abs/hep-th/0412141}{{\tt hep-th/0412141}}].

\bibitem{Sakai:2005yt}
T.~Sakai and S.~Sugimoto, {\it {More on a holographic dual of QCD}},  {\em
  Prog. Theor. Phys.} {\bf 114} (2005) 1083--1118,
  [\href{http://arxiv.org/abs/hep-th/0507073}{{\tt hep-th/0507073}}].

\bibitem{Maldacena:1997re}
J.~M. Maldacena, {\it {The Large N limit of superconformal field theories and
  supergravity}},  {\em Adv.Theor.Math.Phys.} {\bf 2} (1998) 231--252,
  [\href{http://arxiv.org/abs/hep-th/9711200}{{\tt hep-th/9711200}}].

\bibitem{Witten:1998zw}
E.~Witten, {\it {Anti-de Sitter space, thermal phase transition, and
  confinement in gauge theories}},  {\em Adv.Theor.Math.Phys.} {\bf 2} (1998)
  505--532, [\href{http://arxiv.org/abs/hep-th/9803131}{{\tt hep-th/9803131}}].

\bibitem{Hata:2007mb}
H.~Hata, T.~Sakai, S.~Sugimoto, and S.~Yamato, {\it {Baryons from instantons in
  holographic QCD}},  {\em Prog. Theor. Phys.} {\bf 117} (2007) 1157,
  [\href{http://arxiv.org/abs/hep-th/0701280}{{\tt hep-th/0701280}}].

\bibitem{Brunner:2018wbv}
F.~Br{\"u}nner, J.~Leutgeb, and A.~Rebhan, {\it {A broad pseudovector glueball
  from holographic QCD}},  {\em Phys. Lett.} {\bf B788} (2019) 431--435,
  [\href{http://arxiv.org/abs/1807.10164}{{\tt arXiv:1807.10164}}].

\bibitem{Leutgeb:2019lqu}
J.~Leutgeb and A.~Rebhan, {\it {Witten-Veneziano mechanism and pseudoscalar
  glueball-meson mixing in holographic QCD}},
  \href{http://arxiv.org/abs/1909.12352}{{\tt arXiv:1909.12352}}.

\bibitem{Bergman:2007wp}
O.~Bergman, G.~Lifschytz, and M.~Lippert, {\it {Holographic Nuclear Physics}},
  {\em JHEP} {\bf 11} (2007) 056, [\href{http://arxiv.org/abs/0708.0326}{{\tt
  arXiv:0708.0326}}].

\bibitem{McLerran:2007qj}
L.~McLerran and R.~D. Pisarski, {\it {Phases of cold, dense quarks at large
  $N_c$}},  {\em Nucl.Phys.} {\bf A796} (2007) 83--100,
  [\href{http://arxiv.org/abs/0706.2191}{{\tt arXiv:0706.2191}}].

\bibitem{Philipsen:2019qqm}
O.~Philipsen and J.~Scheunert, {\it {QCD in the heavy dense regime for general
  $N_c$: On the existence of quarkyonic matter}},  {\em JHEP} {\bf 11} (2019)
  022, [\href{http://arxiv.org/abs/1908.03136}{{\tt arXiv:1908.03136}}].

\bibitem{Bigazzi:2014qsa}
F.~Bigazzi and A.~L. Cotrone, {\it {Holographic QCD with Dynamical Flavors}},
  {\em JHEP} {\bf 01} (2015) 104, [\href{http://arxiv.org/abs/1410.2443}{{\tt
  arXiv:1410.2443}}].

\bibitem{Li:2016gtz}
S.-w. Li and T.~Jia, {\it {Dynamically flavored description of holographic QCD
  in the presence of a magnetic field}},  {\em Phys. Rev.} {\bf D96} (2017),
  no.~6 066032, [\href{http://arxiv.org/abs/1604.07197}{{\tt
  arXiv:1604.07197}}].

\bibitem{Horigome:2006xu}
N.~Horigome and Y.~Tanii, {\it {Holographic chiral phase transition with
  chemical potential}},  {\em JHEP} {\bf 01} (2007) 072,
  [\href{http://arxiv.org/abs/hep-th/0608198}{{\tt hep-th/0608198}}].

\bibitem{Li:2015uea}
S.-w. Li, A.~Schmitt, and Q.~Wang, {\it {From holography towards real-world
  nuclear matter}},  {\em Phys. Rev.} {\bf D92} (2015), no.~2 026006,
  [\href{http://arxiv.org/abs/1505.04886}{{\tt arXiv:1505.04886}}].

\bibitem{Nambu:1961tp}
Y.~Nambu and G.~Jona-Lasinio, {\it {Dynamical Model of Elementary Particles
  Based on an Analogy with Superconductivity. 1.}},  {\em Phys. Rev.} {\bf 122}
  (1961) 345--358.

\bibitem{Nambu:1961fr}
Y.~Nambu and G.~Jona-Lasinio, {\it {Dynamical model of elementary particles
  based on an analogy with superconductivity. II}},  {\em Phys.Rev.} {\bf 124}
  (1961) 246--254.

\bibitem{Preis:2012fh}
F.~Preis, A.~Rebhan, and A.~Schmitt, {\it {Inverse magnetic catalysis in field
  theory and gauge-gravity duality}},  {\em Lect.Notes Phys.} {\bf 871} (2013)
  51--86, [\href{http://arxiv.org/abs/1208.0536}{{\tt arXiv:1208.0536}}].

\bibitem{Preis:2010cq}
F.~Preis, A.~Rebhan, and A.~Schmitt, {\it {Inverse magnetic catalysis in dense
  holographic matter}},  {\em JHEP} {\bf 1103} (2011) 033,
  [\href{http://arxiv.org/abs/1012.4785}{{\tt arXiv:1012.4785}}].

\bibitem{BitaghsirFadafan:2018uzs}
K.~Bitaghsir~Fadafan, F.~Kazemian, and A.~Schmitt, {\it {Towards a holographic
  quark-hadron continuity}},  {\em JHEP} {\bf 03} (2019) 183,
  [\href{http://arxiv.org/abs/1811.08698}{{\tt arXiv:1811.08698}}].

\bibitem{Bergman:2007pm}
O.~Bergman, S.~Seki, and J.~Sonnenschein, {\it {Quark mass and condensate in
  HQCD}},  {\em JHEP} {\bf 12} (2007) 037,
  [\href{http://arxiv.org/abs/0708.2839}{{\tt arXiv:0708.2839}}].

\bibitem{Dhar:2007bz}
A.~Dhar and P.~Nag, {\it {Sakai-Sugimoto model, Tachyon Condensation and Chiral
  symmetry Breaking}},  {\em JHEP} {\bf 01} (2008) 055,
  [\href{http://arxiv.org/abs/0708.3233}{{\tt arXiv:0708.3233}}].

\bibitem{Dhar:2008um}
A.~Dhar and P.~Nag, {\it {Tachyon condensation and quark mass in modified
  Sakai-Sugimoto model}},  {\em Phys. Rev.} {\bf D78} (2008) 066021,
  [\href{http://arxiv.org/abs/0804.4807}{{\tt arXiv:0804.4807}}].

\bibitem{Aharony:2008an}
O.~Aharony and D.~Kutasov, {\it {Holographic Duals of Long Open Strings}},
  {\em Phys. Rev.} {\bf D78} (2008) 026005,
  [\href{http://arxiv.org/abs/0803.3547}{{\tt arXiv:0803.3547}}].

\bibitem{Hashimoto:2008sr}
K.~Hashimoto, T.~Hirayama, F.-L. Lin, and H.-U. Yee, {\it {Quark Mass
  Deformation of Holographic Massless QCD}},  {\em JHEP} {\bf 07} (2008) 089,
  [\href{http://arxiv.org/abs/0803.4192}{{\tt arXiv:0803.4192}}].

\bibitem{McNees:2008km}
R.~McNees, R.~C. Myers, and A.~Sinha, {\it {On quark masses in holographic
  QCD}},  {\em JHEP} {\bf 11} (2008) 056,
  [\href{http://arxiv.org/abs/0807.5127}{{\tt arXiv:0807.5127}}].

\bibitem{Argyres:2008sw}
P.~C. Argyres, M.~Edalati, R.~G. Leigh, and J.~F. Vazquez-Poritz, {\it {Open
  Wilson Lines and Chiral Condensates in Thermal Holographic QCD}},  {\em Phys.
  Rev.} {\bf D79} (2009) 045022, [\href{http://arxiv.org/abs/0811.4617}{{\tt
  arXiv:0811.4617}}].

\bibitem{Hashimoto:2009hj}
K.~Hashimoto, T.~Hirayama, and D.~K. Hong, {\it {Quark Mass Dependence of
  Hadron Spectrum in Holographic QCD}},  {\em Phys. Rev.} {\bf D81} (2010)
  045016, [\href{http://arxiv.org/abs/0906.0402}{{\tt arXiv:0906.0402}}].

\bibitem{Seki:2012tt}
S.~Seki and S.-J. Sin, {\it {Chiral Condensate in Holographic QCD with Baryon
  Density}},  {\em JHEP} {\bf 08} (2012) 009,
  [\href{http://arxiv.org/abs/1206.5897}{{\tt arXiv:1206.5897}}].

\bibitem{Kobayashi:2006sb}
S.~Kobayashi, D.~Mateos, S.~Matsuura, R.~C. Myers, and R.~M. Thomson, {\it
  {Holographic phase transitions at finite baryon density}},  {\em JHEP} {\bf
  02} (2007) 016, [\href{http://arxiv.org/abs/hep-th/0611099}{{\tt
  hep-th/0611099}}].

\bibitem{Evans:2010iy}
N.~Evans, A.~Gebauer, K.-Y. Kim, and M.~Magou, {\it {Holographic Description of
  the Phase Diagram of a Chiral Symmetry Breaking Gauge Theory}},  {\em JHEP}
  {\bf 03} (2010) 132, [\href{http://arxiv.org/abs/1002.1885}{{\tt
  arXiv:1002.1885}}].

\bibitem{Evans:2011mu}
N.~Evans, A.~Gebauer, and K.-Y. Kim, {\it {E, B, $\mu$, T Phase Structure of
  the D3/D7 Holographic Dual}},  {\em JHEP} {\bf 05} (2011) 067,
  [\href{http://arxiv.org/abs/1103.5627}{{\tt arXiv:1103.5627}}].

\bibitem{Evans:2011eu}
N.~Evans, A.~Gebauer, M.~Magou, and K.-Y. Kim, {\it {Towards a Holographic
  Model of the QCD Phase Diagram}},  {\em J. Phys.} {\bf G39} (2012) 054005,
  [\href{http://arxiv.org/abs/1109.2633}{{\tt arXiv:1109.2633}}].

\bibitem{Jarvinen:2011qe}
M.~J{\"a}rvinen and E.~Kiritsis, {\it {Holographic Models for QCD in the
  Veneziano Limit}},  {\em JHEP} {\bf 03} (2012) 002,
  [\href{http://arxiv.org/abs/1112.1261}{{\tt arXiv:1112.1261}}].

\bibitem{Ishii:2019gta}
T.~Ishii, M.~J{\"a}rvinen, and G.~Nijs, {\it {Cool baryon and quark matter in
  holographic QCD}},  {\em JHEP} {\bf 07} (2019) 003,
  [\href{http://arxiv.org/abs/1903.06169}{{\tt arXiv:1903.06169}}].

\bibitem{Gursoy:2017wzz}
U.~G{\"u}rsoy, M.~J{\"a}rvinen, and G.~Nijs, {\it {Holographic QCD in the
  Veneziano Limit at a Finite Magnetic Field and Chemical Potential}},  {\em
  Phys. Rev. Lett.} {\bf 120} (2018), no.~24 242002,
  [\href{http://arxiv.org/abs/1707.00872}{{\tt arXiv:1707.00872}}].

\bibitem{Fromm:2011qi}
M.~Fromm, J.~Langelage, S.~Lottini, and O.~Philipsen, {\it {The QCD
  deconfinement transition for heavy quarks and all baryon chemical
  potentials}},  {\em JHEP} {\bf 01} (2012) 042,
  [\href{http://arxiv.org/abs/1111.4953}{{\tt arXiv:1111.4953}}].

\bibitem{Rebhan:2014rxa}
A.~Rebhan, {\it {The Witten-Sakai-Sugimoto model: A brief review and some
  recent results}},  {\em EPJ Web Conf.} {\bf 95} (2015) 02005,
  [\href{http://arxiv.org/abs/1410.8858}{{\tt arXiv:1410.8858}}].

\bibitem{Mandal:2011ws}
G.~Mandal and T.~Morita, {\it {Gregory-Laflamme as the
  confinement/deconfinement transition in holographic QCD}},  {\em JHEP} {\bf
  09} (2011) 073, [\href{http://arxiv.org/abs/1107.4048}{{\tt
  arXiv:1107.4048}}].

\bibitem{Preis:2016fsp}
F.~Preis and A.~Schmitt, {\it {Layers of deformed instantons in holographic
  baryonic matter}},  {\em JHEP} {\bf 07} (2016) 001,
  [\href{http://arxiv.org/abs/1606.00675}{{\tt arXiv:1606.00675}}].

\bibitem{Aharony:2006da}
O.~Aharony, J.~Sonnenschein, and S.~Yankielowicz, {\it {A Holographic model of
  deconfinement and chiral symmetry restoration}},  {\em Annals Phys.} {\bf
  322} (2007) 1420--1443, [\href{http://arxiv.org/abs/hep-th/0604161}{{\tt
  hep-th/0604161}}].

\bibitem{Antonyan:2006vw}
E.~Antonyan, J.~A. Harvey, S.~Jensen, and D.~Kutasov, {\it {NJL and QCD from
  string theory}},  \href{http://arxiv.org/abs/hep-th/0604017}{{\tt
  hep-th/0604017}}.

\bibitem{Davis:2007ka}
J.~L. Davis, M.~Gutperle, P.~Kraus, and I.~Sachs, {\it {Stringy NJL and
  Gross-Neveu models at finite density and temperature}},  {\em JHEP} {\bf
  0710} (2007) 049, [\href{http://arxiv.org/abs/0708.0589}{{\tt
  arXiv:0708.0589}}].

\bibitem{GellMann:1968rz}
M.~Gell-Mann, R.~J. Oakes, and B.~Renner, {\it {Behavior of current divergences
  under SU(3) x SU(3)}},  {\em Phys. Rev.} {\bf 175} (1968) 2195--2199.

\bibitem{Brunner:2015oga}
F.~Br{\"u}nner and A.~Rebhan, {\it {Constraints on the $\eta \eta'$ decay rate
  of a scalar glueball from gauge/gravity duality}},  {\em Phys. Rev.} {\bf
  D92} (2015), no.~12 121902, [\href{http://arxiv.org/abs/1510.07605}{{\tt
  arXiv:1510.07605}}].

\bibitem{Hashimoto:2009st}
K.~Hashimoto, N.~Iizuka, T.~Ishii, and D.~Kadoh, {\it {Three-flavor quark mass
  dependence of baryon spectra in holographic QCD}},  {\em Phys. Lett.} {\bf
  B691} (2010) 65--71, [\href{http://arxiv.org/abs/0910.1179}{{\tt
  arXiv:0910.1179}}].

\bibitem{Bigazzi:2018cpg}
F.~Bigazzi and P.~Niro, {\it {Neutron-proton mass difference from gauge/gravity
  duality}},  {\em Phys. Rev.} {\bf D98} (2018), no.~4 046004,
  [\href{http://arxiv.org/abs/1803.05202}{{\tt arXiv:1803.05202}}].

\bibitem{Ewerz:2016zsx}
C.~Ewerz, O.~Kaczmarek, and A.~Samberg, {\it {Free Energy of a Heavy
  Quark-Antiquark Pair in a Thermal Medium from AdS/CFT}},  {\em JHEP} {\bf 03}
  (2018) 088, [\href{http://arxiv.org/abs/1605.07181}{{\tt arXiv:1605.07181}}].

\bibitem{Bak:2007fk}
D.~Bak, A.~Karch, and L.~G. Yaffe, {\it {Debye screening in strongly coupled
  N=4 supersymmetric Yang-Mills plasma}},  {\em JHEP} {\bf 08} (2007) 049,
  [\href{http://arxiv.org/abs/0705.0994}{{\tt arXiv:0705.0994}}].

\bibitem{Mateos:2007vc}
D.~Mateos, S.~Matsuura, R.~C. Myers, and R.~M. Thomson, {\it {Holographic phase
  transitions at finite chemical potential}},  {\em JHEP} {\bf 11} (2007) 085,
  [\href{http://arxiv.org/abs/0709.1225}{{\tt arXiv:0709.1225}}].

\bibitem{Callan:1997kz}
C.~G. Callan and J.~M. Maldacena, {\it {Brane death and dynamics from the
  Born-Infeld action}},  {\em Nucl. Phys.} {\bf B513} (1998) 198--212,
  [\href{http://arxiv.org/abs/hep-th/9708147}{{\tt hep-th/9708147}}].

\bibitem{Gibbons:1997xz}
G.~W. Gibbons, {\it {Born-Infeld particles and Dirichlet p-branes}},  {\em
  Nucl. Phys.} {\bf B514} (1998) 603--639,
  [\href{http://arxiv.org/abs/hep-th/9709027}{{\tt hep-th/9709027}}].

\bibitem{Evans:2007jr}
N.~Evans and E.~Threlfall, {\it {Quark Mass in the Sakai-Sugimoto Model of
  Chiral Symmetry Breaking}},  \href{http://arxiv.org/abs/0706.3285}{{\tt
  arXiv:0706.3285}}.

\bibitem{Bhattacharyya:2012rp}
A.~Bhattacharyya, P.~Deb, S.~K. Ghosh, R.~Ray, and S.~Sur, {\it {Thermodynamic
  Properties of Strongly Interacting Matter in Finite Volume using
  Polyakov-Nambu-Jona-Lasinio Model}},  {\em Phys. Rev.} {\bf D87} (2013),
  no.~5 054009, [\href{http://arxiv.org/abs/1212.5893}{{\tt arXiv:1212.5893}}].

\bibitem{Marty:2013ita}
R.~Marty, E.~Bratkovskaya, W.~Cassing, J.~Aichelin, and H.~Berrehrah, {\it
  {Transport coefficients from the Nambu-Jona-Lasinio model for $SU(3)_f$}},
  {\em Phys. Rev.} {\bf C88} (2013) 045204,
  [\href{http://arxiv.org/abs/1305.7180}{{\tt arXiv:1305.7180}}].

\end{thebibliography}\endgroup

\end{document}